\documentclass[12pt]{article}
\usepackage{scipp}  \usepackage{equations}
\usepackage{psfig}

\begin{document}
%
\pagestyle{empty}
\topmargin-2pc
\begin{flushright}
SCIPP 97/11     \\
hep--ph/9707213
\end{flushright}
\vskip4cm

\begin{center}
{\Large\bf  Higgs Boson Masses and Couplings \\[6pt]
 in the Minimal Supersymmetric Model}\\[1cm]
{\large Howard E. Haber}\\[3pt]
{\it Santa Cruz Institute for Particle Physics  \\
   University of California, Santa Cruz, CA 95064, U.S.A.} \\[1.5cm]

{\bf Abstract}
\end{center}

The Higgs sector of the Minimal Supersymmetric Model (MSSM) is a
CP-conserving two-Higgs doublet model that depends, at tree-level,
on two Higgs sector parameters.  In order to accurately determine the
phenomenological implications of this model, one must include
the effects of radiative corrections.
The leading contributions to the one-loop radiative corrections are
exhibited;  large logarithms are resummed by the renormalization group
method.  Implications for Higgs phenomenology are briefly discussed.

\vskip3cm
\centerline{To appear in {\it  Perspectives on Higgs Physics II}}
\centerline{Gordon L. Kane, editor (World Scientific, Singapore, 1997)}
\vfill
\clearpage
\def\solid{(---------)}
\def\dashes{($-~-~-~-$)}
\def\dotdot{($\cdot~\cdot~\cdot~\cdot~\cdot~\cdot\,$)}
\def\daashdash{(-----~$-$~-----~$-$)}
\def\dotdash{($\cdot~-~\cdot~-$)}
\def\dotdotdash{($\cdot~\cdot~-~\cdot~\cdot~-$)}
\def\dotdotdotdash{($\cdot~\cdot~\cdot~-$)}
\def\dotdotdotdotdash{($\cdot~\cdot~\cdot~\cdot~-$)}
\def\dotdashdotdotdash{($\cdot~-~\cdot~\cdot~-$)}
\def\dotdotdotdashdash{($\cdot~\cdot~\cdot~-~-$)}
\def\dotdashdash{($\cdot~-~-~\cdot~-~-$)}
\def\dotdotdashdash{($\cdot~\cdot~-~-$)}
\def\ee#1{\times 10^{{#1}}}
\def\gev{~{\rm GeV}}
\def\tev{~{\rm TeV}}
\def\pbi{~{\rm pb}^{-1}}
\def\fbi{~{\rm fb}^{-1}}
\def\fb{~{\rm fb}}
\def\pb{~{\rm pb}}
\def\introsection{1}
\def\smsection{2}
\def\searchsection{3}
\def\susysection{4}
\def\beyondsection{5}
\def\othersection{6}
\def\alternativessection{7}
\def\finalsection{8}
\def\lncj#1{{\it Lett. Nuovo Cim.} {\bf{#1}}}
\def\prdj#1{{\it Phys. Rev.} {\bf D{#1}}}
\def\npbj#1{{\it Nucl. Phys.} {\bf B{#1}}}
\def\prlj#1{{\it Phys. Rev. Lett.} {\bf {#1}}}
\def\plbj#1{{\it Phys. Lett.} {\bf B{#1}}}
\def\zpcj#1{{\it Z. Phys.} {\bf C{#1}}}
\def\rmpj#1{{\it Rev. Mod. Phys.} {\bf {#1}}}
\def\prepj#1{{\it Phys. Rep.} {\bf {#1}}}
\def\ibid{{\it ibid.}}
\def\lamqcd{\Lambda_{QCD}}
\def\lqcd{\Lambda_{QCD}}
\def\wtilde{\widetilde}
\def\wtil{\widetilde}
\def\til{\tilde}
\def\etal{{\it et al.}}
\def\etc{{\it etc.}}
\def\ls#1{\ifmath{_{\lower1.5pt\hbox{$\scriptstyle #1$}}}}
\def\qkh{\xi}
\def\square{\boxxit{0.4pt}{\fillboxx{7pt}{7pt}}\hskip-0.4pt}
    \def\boxxit#1#2{\vbox{\hrule height #1 \hbox {\vrule width #1
             \vbox{#2}\vrule width #1 }\hrule height #1 } }
    \def\fillboxx#1#2{\hbox to #1{\vbox to #2{\vfil}\hfil}
    }
\def\bdbdb{B_d-\overline B_d}
\def\mww{M_{WW}}
\def\mzz{M_{ZZ}}
\def\mjj{M_{jj}}
\def\mplanck{m_{Pl}}
\def\Lam{\Lambda}
\def\gt{h}
\def\hdag{^{\dag}}
\def\gpl{G^+}
\def\gm{G^-}
\def\gn{G^0}
\def\gpm{G^{\pm}}
\def\gmp{G^{\mp}}
\def\monetwo{m_{12}}
\def\monetwos{\monetwo^2}
\def\pha{\phi_1}
\def\phb{\phi_2}
\def\phad{\phi_1^{\dag}}
\def\phbd{\phi_2^{\dag}}
\def\phaa{\phad\pha}
\def\phbb{\phbd\phb}
\def\phab{\phad\phb}
\def\phba{\phbd\pha}
\def\rts{\sqrt{s}}
\def\ptmiss{p_T^{\,miss}}
\def\ptmisssq{(p_T^{\,miss})^2}
\def\etmiss{E_T^{\,miss}}
\def\zsm{Z}
\def\mzsm{m_{\zsm}}
\def\wpm{W^{\pm}}
\def\mwpm{m_{\wpm}}
\def\mt{m_t}
\def\mb{m_b}
\def\mc{m_c}
\def\mtoponium{m_{\Theta}}
\def\sutwo{SU(2)}
\def\ss{\scriptscriptstyle}
\def\sm{Standard Model}
\def\dexotic{D}
\def\u{\undertext}
\def\p{\phi}
\def\hsm{\phi^0}
\def\mhsm{m_{\hsm}}
\def\zone{Z}
\def\mf{m_f}
\def\amuup{A^\mu}
\def\mh{m_h}
\def\hl{h^0}
\def\mhl{m_{\hl}}
\def\hh{H^0}
\def\mhh{m_{\hh}}
\def\hpm{H^{\pm}}
\def\mhpm{m_{\hpm}}

\def\madii{{\it Proceedings of the 1987 Madison Workshop on
``From Colliders to Supercolliders''}, edited by V. Barger and F.
Halzen}
\def\berkii{{\it Proceedings of the 1987 Berkeley Workshop on
``Experiments, Detectors and
Experimental Areas for the Supercollider''}, edited by R. Donaldson
and M. Gilchriese}
\def\smii{{\it Proceedings of the 1984 Snowmass Workshop on the
Design and Utilization of the Superconducting Super
Collider, \rm edited by R.~Donaldson and J.~G.~Morfin}}
\def\smiii{{\it Proceedings of the 1986 Snowmass Workshop on ``Physics
of the Superconducting Super
Collider'', \rm edited by R.~Donaldson and J.~Marx}}
\def\smiv{{\it Proceedings of the 1988 Snowmass Workshop on ``High Energy
Physics in the 1990's''}}
\def\ucla{{\it Proceedings of the 1986 UCLA Workshop on Observable
Standard Model Physics at the SSC}, \rm edited by H-U Bengtsson,
C. Buchanan, T. Gottschalk, and A. Soni}
\def\lath{{\it Proceedings of the Workshop on Physics at Future
Colliders}, La Thuile (1987)}
\def\lhchilum{{\it The Feasibility of Experiments at High Luminosity
at the Large Hadron Collider}, ed. by J.H. Mulvey, CERN-88-02 (1988) }
\def\ericei{{\it INFN Eloisatron Project Working Group Report}, ed.
by A. Ali, Erice-Trapani (1988)}
\def\ucdwkshop{{\it Proceedings of the 1988 U.C. Davis Workshop on
Intermediate Mass and Non-Minimal Higgs Bosons}, \rm edited by
J.F. Gunion and L. Roszkowski}
\def\munichhep{{\it Proceedings of the XXIV International Conference
on High Energy Physics}, (Munich, 1988)}
\def\amesconf{{\it Proceedings of `Beyond the Standard Model'}, (Ames,
1988)}
 
\def\tho{\theta^0}
\def\rttwo{\sqrt{2}}
\def\vev{v.e.v.}
\def\kap{\kappa}
\def\kappr{\kappa^{\prime}}
\def\kapprs{\kappa^{\prime\,2}}
\def\dr{\Delta_R}
\def\dl{\Delta_L}
\def\dro{\dr^0}
\def\drp{\dr^+}
\def\drpp{\dr^{++}}
\def\dlo{\dl^0}
\def\dlp{\dl^+}
\def\dlm{\dl^-}
\def\dlpp{\dl^{++}}
\def\dlmm{\dl^{--}}
\def\sutwotwoone{SU(2)_L\times SU(2)_R \times U(1)_{B-L}}
\def\sutwo{SU(2)}
\def\uonebl{U(1)_{B-L}}
\def\zone{Z_1}
\def\ztwo{Z_2}
\def\wone{W_1}
\def\wtwo{W_2}
\def\mzone{m_{\zone}}
\def\mztwo{m_{\ztwo}}
\def\mwone{m_{\wone}}
\def\mwtwo{m_{\wtwo}}
\def\lr{LR}
\def\ssc{SSC}
\def\bb{\bar b}
\def\tb{\bar t}
\def\cosa{\cos\alpha}
\def\sina{\sin\alpha}
\def\tana{\tan\alpha}
\def\cosbma{\cos(\beta-\alpha)}
\def\sinbma{\sin(\beta-\alpha)}
\def\cosbpa{\cos(\beta+\alpha)}
\def\sinbpa{\sin(\beta+\alpha)}
\def\ifmath#1{\relax\ifmmode #1\else $#1$\fi}
\def\half{\ifmath{{\textstyle{1 \over 2}}}}
\def\threeeighths{\ifmath{{\textstyle{3 \over 8}}}}
\def\fivehalfs{\ifmath{{\textstyle{5 \over 2}}}}
\def\quarter{\ifmath{{\textstyle{1 \over 4}}}}
\def\3quarter{{\textstyle{3 \over 4}}}
\def\third{\ifmath{{\textstyle{1 \over 3}}}}
\def\twothirds{{\textstyle{2 \over 3}}}
\def\fourthirds{{\textstyle{4 \over 3}}}
\def\eightthirds{{\textstyle{8 \over 3}}}
\def\fourth{\ifmath{{\textstyle{1\over 4}}}}
\def\nicefrac#1#2{\hbox{${#1\over #2}$}}
\def\vsqsum{(v_1^2+v_2^2)}
\def\gpr{g^{\prime}}
\def\gprs{g^{\prime\, 2}}
\def\twogam{\gamma\gamma}
\def\smod{SM}
\def\susy{SUSY}
\def\ptb{p_T^{back}}
\def\tauptaum{\tau^+\tau^-}
\def\mupmum{\mu^+\mu^-}
\def\qpr{q^{\prime}}
\def\qp{q^{\prime}}
\def\qqpbar{q \bar \qp}
\def\db{\delta_B}
\def\h{h}
\def\mh{m_{\h}}
\def\lplm{l^+l^-}
\def\eps{\epsilon}
\def\tanb{\tan\beta}
\def\cotb{\cot\beta}
\def\sinb{\sin\beta}
\def\cosb{\cos\beta}
\def\cota{\cot\alpha}
\def\tana{\tan\alpha}
\def\sw{s_W}
\def\cw{c_W}
\def\sb{s_{\beta}}
\def\cb{c_{\beta}}
\def\thw{\theta_W}
\def\tanw{\tan\thw}
\def\mpr{M^{\prime}}
\def\mprs{M^{\prime \, 2}}
\def\mw{m_W}
\def\mz{m_Z}
\def\hpm{H^{\pm}}
\def\mhpm{m_{\hpm}}
\def\mhp{m_{\hp}}
\def\hp{H^+}
\def\hm{H^-}
\def\hone{H^0_1}
\def\htwo{H^0_2}
\def\hthree{H^0_3}
\def\mhone{m_{\hone}}
\def\mhtwo{m_{\htwo}}
\def\mhthree{m_{\hthree}}
\def\rta{\rightarrow}
\def\lra{\leftrightarrow}
\def\sq{\widetilde q}
\def\sqb{\overline{\sq}}
\def\slep{\widetilde l}
\def\snu{\widetilde \nu}
\def\slb{\overline {\slep}}
\def\gl{\widetilde g}
\def\chitil{\widetilde\chi}
\def\mchipa{M_{\tilde \chi^+_1}}
\def\mchipb{M_{\tilde \chi^+_2}}
\def\mchiza{M_{\tilde \chi^0_1}}
\def\mchizb{M_{\tilde \chi^0_2}}
\def\mchizc{M_{\tilde \chi^0_3}}
\def\mchizd{M_{\tilde \chi^0_4}}
\def\cnone{\chitil^0_1}
\def\cntwo{\chitil^0_2}
\def\cnthree{\chitil^0_3}
\def\cnfour{\chitil^0_4}
\def\cnsum{\chitil^0_{2,3,4}}
\def\ccone{\chitil^{\pm}_1}
\def\cctwo{\chitil^{\pm}_2}
\def\ccsum{\chitil^{\pm}_{1,2}}
\def\mcnone{m_{\chitil^0_1}}
\def\mcntwo{m_{\chitil^0_2}}
\def\mcnthree{m_{\chitil^0_3}}
\def\mcnfour{m_{\chitil^0_4}}
\def\mccone{m_{\chitil^+_1}}
\def\mcctwo{m_{\chitil^+_2}}
\def\hmp{H^{\mp}}
\def\wmp{W^{\mp}}
\def\wpm{W^{\pm}}
\def\wp{W^+}
\def\wm{W^-}
\def\sql{\widetilde q_L}
\def\sqlb{\overline{\widetilde q}_L}
\def\sqr{\widetilde q_R}
\def\sqrb{\overline{\widetilde q}_R}
\def\sfl{\widetilde f_L}
\def\sflb{\overline{\widetilde f}_L}
\def\sfr{\widetilde f_R}
\def\sfrb{\overline{\widetilde f}_R}
\def\sul{\widetilde u_L}
\def\sur{\widetilde u_R}
\def\sdl{\widetilde d_L}
\def\sdr{\widetilde d_R}
\def\sulb{\overline{\widetilde u}_L}
\def\surb{\overline{\widetilde u}_R}
\def\sdlb{\overline{\widetilde d}_L}
\def\sdrb{\overline{\widetilde d}_R}
\def\stl{\widetilde t_L}
\def\str{\widetilde t_R}
\def\sbl{\widetilde b_L}
\def\sbr{\widetilde b_R}
\def\stlb{\overline{\widetilde t}_L}
\def\strb{\overline{\widetilde t}_R}
\def\sblb{\overline{\widetilde b}_L}
\def\sbrb{\overline{\widetilde b}_R}
\def\msul{M_{\tilde u_L}}
\def\msur{M_{\tilde u_R}}
\def\msdl{M_{\tilde d_L}}
\def\msdr{M_{\tilde d_R}}
\def\msq{M_{\tilde q}}
\def\msf{M_{\tilde f}}
\def\mi{M_i}
\def\hl{h^0}
\def\hh{H^0}
\def\ha{A^0}
\def\mhl{m_{\hl}}
\def\mhh{m_{\hh}}
\def\mha{m_{\ha}}
\def\eps{\epsilon}
\def\sutwoone{SU(2)\times U(1)}
\def\sutwo{SU(2)}
\def\lam{\lambda}
\def\lampr{\lambda^{\prime}}
\def\gam{\gamma}
\def\rrat{R_{W \h}}
\def\rrati{R_{W \hh}}
\def\rratii{R_{W \hl}}
\def\rrrat{R_{W \h \gamma}}
\def\rrratii{R_{W \hl \gamma}}
 
\def\SLAC{\centerline{\it Stanford Linear Accelerator Center}
\centerline{\it Stanford University, Stanford, CA 94305}}
\def\AUSTIN{\centerline{\it Center for Particle Theory}
\centerline{\it University of Texas at Austin, Austin, TX 78712}}
\def\MICH{\centerline{\it Randall Physics Laboratory}
\centerline{\it University of Michigan, Ann Arbor, MI 48109}}
\def\IIT{\centerline{\it Department of Physics}
\centerline{\it Illinois Institute of Technology, Chicago, IL 60616}}
\def\LBL{\centerline{\it Lawrence Berkeley Laboratory,
Berkeley CA 94720}}
\def\BHVN{  \centerline{\it Physics Department}
\centerline{\it Brookhaven National Laboratory,
 Upton, Long Island, NY 11973}   }
\def\SCIPP{\centerline {\it Santa Cruz Institute for Particle Physics}
  \centerline{\it University of California, Santa Cruz, CA 95064}}
\def\DAVIS{\centerline {\it Department of Physics}
  \centerline{\it University of California, Davis, CA 95616}}
\def\WARSAW{\centerline {\it Institute of Theoretical Physics}
  \centerline {\it Warsaw University, Warsaw, Poland}}
\def\doeack{\foot{Work supported, in part, by the Department of Energy.}}
\def\ictp{\foot{To appear in {\it Proceedings of the 1987 Conference on
``Search for Scalar Particles: Experimental and Theoretical Aspects''},
International Centre for Theoretical Physics (July, 1987),
eds. J.C. Pati and Q. Shafi.}}
 
\def\lamfour{\lambda_{min}^{[40]}}
\def\gam{\gamma}
\def\vev{v.e.v.}
\def\Vev{\langle V\rangle}
\def\VEV#1{\langle #1 \rangle}
\def\stb{s_{2\beta}}
\def\ctb{c_{2\beta}}
\def\halph{H_{\alpha}}
\def\hb{H_{\beta}}
\def\hc{H_{\gamma}}
\def\hbf{{\bf H}}
\def\hba{{\bf H}_{\alpha}}
\def\ghww{g_{\halph \wp \wm}}
\def\ghzz{g_{\halph \zone \zone}}
\def\ghzzp{g_{\halph \zone \ztwo}}
\def\ghzpzp{g_{\halph \ztwo \ztwo}}
\def\tanb{\tan\beta}
\def\sinb{\sin\beta}
\def\cosb{\cos\beta}
\def\sb{s_{\beta}}
\def\cb{c_{\beta}}
\def\vp{V^{\prime}}
\def\cvvp{C_{V\vp}}
\def\chitil{\widetilde \chi}
\def\lsp{\chitil^0_1}
\def\lcsp{\chitil^+_1}
\def\gp{g^{\prime}}
\def\mpr{M^{\prime}}
\def\gl{\widetilde g}
\def\mgl{M_{\gl}}
\def\thmix{\theta_{mix}}
\def\del{\delta}
\def\wpm{W^{\pm}}
\def\wmp{W^{\mp}}
\def\wp{W^+}
\def\wm{W^-}
\def\mw{m_W}
\def\mz{m_{Z}}
\def\mzsq{\mz^2}
\def\lam{\lambda}
\def\lama{\lambda A}
\def\vone{v_1}
\def\vtwo{v_2}
\def\qqbar{q \bar q}
\def\hi{H_1}
\def\hid{\hi^{\dag}}
\def\hii{H_2}
\def\hiid{\hii^{\dag}}
\def\none{N_1} \def\ntwo{N_2}
\def\noned{\none^{\dag}} \def\ntwod{\ntwo^{\dag}}
\def\hone{H_1^0}
\def\htwo{h^0}
\def\hthree{A^0}
\def\hzero{H_0^0}
\def\nzero{N^0}
\def\hpm{H^{\pm}}
\def\hmp{H^{\mp}}
\def\hp{H^+}
\def\hm{H^-}
\def\mhone{m_{\hone}}
\def\mhtwo{m_{\htwo}}
\def\mhzero{m_{\hzero}}
\def\mhpm{m_{\hpm}}
\def\mhthree{m_{\hthree}}
\def\zone{Z_1}
\def\mzone{m_{\zone}}
\def\ztwo{Z_2}
\def\mztwo{m_{\ztwo}}
\def\epem{e^+e^-}
\def\mupmum{\mu ^+ \mu ^-}
\def\tauplmi{\tau^+ \tau^-}
\def\lplm{l^+ l^-}
\def\thw{\theta _W}
\def\suu{SU(2)_L \times U(1)}
\def\esix{E_6}
\def\legroup{SU(2)_L\times U(1)_Y\times U(1)_{Y^{\prime}}}
\def\pitwo{\coeff \pi 2}
\def\pifour{\coeff \pi 4}
\def\hztwo{H^0_{\ztwo}}
\def\mhztwo{m_{\hztwo}}
\def\hdeg{H^0_{deg}}
\def\mhdeg{m_{\hdeg}}
\def\rta{\rightarrow}
\def\ghw{g_{\ztwo\hpm\wmp}}
\def\ghz{g_{\ztwo\hdeg\zone}}
\def\ghlz{g_{\ztwo\hl\zone}}
\def\fhw{f_{\hpm\wmp}}
\def\fhz{f_{\hdeg\zone}}
\def\fhlz{f_{\hl\zone}}
\def\bee{B^{\ztwo}_{\epem}}
\def\bzone{B^{\zone}_{\lplm}}
\def\bw{B^{W}_{l \nu}}
\def\bratio{B^{\ztwo}_{HV}}
\def\mzh{M_{\zone H}}
\def\fhv{f_{HV}}
\def\cni{\chitil^0_i}
\def\cci{\chitil^{\pm}_i}
\def\cpi{\chitil^+_i}
\def\cmi{\chitil^-_i}
\def\cnj{\chitil^0_j}
\def\ccj{\chitil^{\pm}_j}
\def\cpj{\chitil^+_j}
\def\cmj{\chitil^-_j}
\def\cn{\chitil^0}
\def\cc{\chitil^{\pm}}
\def\cp{\chitil^+}
\def\cm{\chitil^-}
\def\cnone{\chitil^0_1}
\def\cntwo{\chitil^0_2}
\def\cnthree{\chitil^0_3}
\def\cnfour{\chitil^0_4}
\def\cnfive{\chitil^0_5}
\def\cnsix{\chitil^0_6}
\def\cnsum{\chitil^0_{2,3,4}}
\def\ccone{\chitil^{\pm}_1}
\def\cctwo{\chitil^{\pm}_2}
\def\ccsum{\chitil^{\pm}_{1,2}}
\def\cpone{\chitil^+_1}
\def\cptwo{\chitil^+_2}
\def\cnoneb{\overline{\chitil^0_1}}
\def\cntwob{\overline{\chitil^0_2}}
\def\cnthreeb{\overline{\chitil^0_3}}
\def\cnfourb{\overline{\chitil^0_4}}
\def\cnfiveb{\overline{\chitil^0_5}}
\def\cnsixb{\overline{\chitil^0_6}}
\def\cponeb{\overline{\chitil^+_1}}
\def\cptwob{\overline{\chitil^+_2}}
\def\mcnone{m_{\chitil^0_1}}
\def\mcntwo{m_{\chitil^0_2}}
\def\mcnthree{m_{\chitil^0_3}}
\def\mcnfour{m_{\chitil^0_4}}
\def\mccone{m_{\chitil^+_1}}
\def\mcctwo{m_{\chitil^+_2}}
\def\cna{\chitil^0_{\tilde B^{\prime}}}
\def\cnb{\chitil^0_{\tilde W_3}}
\def\cnc{\chitil^0_{\tilde H_1+\tilde H_2}}
\def\cnd{\chitil^0_{\tilde H_1-\tilde H_2}}
\def\cne{\chitil^0_{\tilde N_1+\tilde B_1}}
\def\cnf{\chitil^0_{\tilde N_1-\tilde B_1}}
\def\cca{\chitil^+_{\tilde W^{+}}}
\def\ccb{\chitil^+_{\tilde H^{+}}}
\def\mt{m_t}
\def\mhp{m_{H^+}}
\def\su{SU(3)_C\times SU(2)_L\times U(1)_Y}
\def\ss{\scriptscriptstyle}
\def\sutwo{SU(2)}
\def\sm{Standard Model}
\def\u{\undertext}
\def\p{\phi}
\def\thetaw{\theta_W}
\def\pdeg{\phi^0}
\def\amuup{A^\mu}
\def\hcirc{h^0}
\def\Hcirc{H^0}
\def\Pcirc{P^0}
\def\ppcirc{P^{\prime\, 0}}
\def\mpcirc{M_{\Pcirc}}
\def\zcirc{Z}
\def\mpplus{M_{{P}^+}}
\def\mhplus{M_{{H}^+}}
\def\M2pm{M^2_{P^\pm}}
\def\m2z{\mz^2}
\def\ppm{P^\pm}
\def\etat{\eta_T}

\def\mplj{{\sl Mod. Phys. Lett.}}
\def\phm{\phantom{-}}
\def\mhsm{m_{\phi^0}}
\def\pri{{\, \prime }}
\def\beqno{\begin{eqalignno}}
\def\eeqno{\end{eqalignno}}
\def\beq{\begin{equation}}
\def\eeq{\end{equation}}
\def\eg{{\it e.g.}}
\def\ifmath#1{\relax\ifmmode #1\else $#1$\fi}
\def\calm{{\cal M}}
\def\calu{{\cal U}}
\def\cals{{\cal S}}
\def\calm{{\cal M}}
\def\caln{{\cal N}}
\def\calv{{\cal V}}
\def\call{{\cal L}}
\def\tb  {t_{\beta}}
\def\tbb  {\bar t_{\beta}}
\def\sw  {s_W}
\def\cw  {c_W}
\def\sb  {s_{\beta}}
\def\cb  {c_{\beta}}
\def\stwob  {s_{2\beta}}
\def\ctwob  {c_{2\beta}}
\def\sa  {s_{\alpha}}
\def\ca  {c_{\alpha}}
\def\sab  {s_{\alpha+\beta}}
\def\cab  {c_{\alpha+\beta}}
\def\sba  {s_{\beta-\alpha}}
\def\cba  {c_{\beta-\alpha}}
\def\tanb{\tan\beta}
\def\sinb{\sin\beta}
\def\cosb{\cos\beta}
\def\sina{\sin\alpha}
\def\cosa{\cos\alpha}
\def\cww{c_W^2}
\def\sww{s_W^2}
\def\azzm{A_{ZZ}(m_Z^2)}
\def\ahhm{A_{hh}(m_Z^2)}
\def\deltap{\Delta m^2_h\mid^{apprx}}
\def\deltam{\big(\Delta m^2_h\big)}
\def\factor{{g^2\over 16\pi^2c_W^2}}
\def\factm{{g^2m_Z^2\over 16\pi^2c_W^2}}
\def\factmw{{blablag^2m_Z^2\over 16\pi^2c_W^2}}
\def\sixteenpi{\ifmath{{\textstyle{1 \over16\pi^2}}}}
\def\feightpi{\ifmath{{\textstyle{1 \over48\pi^2}}}}
\def\nmp{M'^2-\mu^2}
\def\nmm{(M^2-\mu^2)^2}
\def\nm{M^2-\mu^2}
\def\nmpp{(M'^2-\mu^2)^2}
\def\lnmp{\ln\big({M'^2\over \mu^2}\big)}
\def\lnm{\ln\big({M^2\over \mu^2}\big)}
\def\lnlr{\ln\big({m_L^2\over m_R^2}\big)}

\def\half{\ifmath{{\textstyle{1 \over 2}}}}
\def\fivehalf{\ifmath{{\textstyle{5 \over 2}}}}
\def\fivesixths{\ifmath{{\textstyle{5 \over 6}}}}
\def\fivetwelfth{\ifmath{{\textstyle{5 \over 12}}}}
\def\thirteenthirds{\ifmath{{\textstyle{13\over 3}}}}
\def\fortythirds{\ifmath{{\textstyle{40 \over 3}}}}
\def\sixtyfourthirds{\ifmath{{\textstyle{64 \over 3}}}}
\def\fourthirds{\ifmath{{\textstyle{4 \over 3}}}}
\def\third{\ifmath{{\textstyle{1 \over 3}}}}
\def\twothirds{\ifmath{{\textstyle{2\over 3}}}}
\def\fourth{\ifmath{{\textstyle{1\over 4}}}}
\def\threefourth{\ifmath{{\textstyle{3\over 4}}}}
\def\fifteenfourth{\ifmath{{\textstyle{15\over 4}}}}
\def\fifth{\ifmath{{\textstyle{1 \over 5}}}}
\def\twelfth{\ifmath{{\textstyle{1 \over 12}}}}
\def\eighth{\ifmath{{\textstyle{1 \over 8}}}}
\def\threeighth{\ifmath{{\textstyle{3 \over 8}}}}
\def\threesixtn{\ifmath{{\textstyle{3 \over16}}}}
\def\eightthirds{\ifmath{{\textstyle{8 \over 3}}}}
\def\thirtytwoninths{\ifmath{{\textstyle{32\over 9}}}}
\def\eightthirds{\ifmath{{\textstyle{8 \over 3}}}}
\def\twentythirds{\ifmath{{\textstyle{20\over 3}}}}
\def\eightninths{\ifmath{{\textstyle{8 \over 9}}}}
\def\twentyninths{\ifmath{{\textstyle{20\over 9}}}}
\def\ninehalf{\ifmath{{\textstyle{9 \over 2}}}}
\def\seventeentwelfth{\ifmath{{\textstyle{17 \over12}}}}
\def\sevennineths{\ifmath{{\textstyle{7\over 9}}}}
\def\thirteennineths{\ifmath{{\textstyle{13\over 9}}}}
\def\sixteenthirds{\ifmath{{\textstyle{16\over 3}}}}
\def\ninefourth{\ifmath{{\textstyle{9 \over 4}}}}
\def\sixth{\ifmath{{\textstyle{1 \over 6}}}}
\def\mzz{m_Z^2}
\def\mweak{M\ls{{\rm weak}}}
\def\msusy{M\ls{{\rm SUSY}}}
\def\msusyy{M\ls{{\rm SUSY}}^2}
\def\half{\ifmath{{\textstyle{1 \over 2}}}}
\def\thalf{\ifmath{{\textstyle{3 \over 2}}}}
\def\fhalf{\ifmath{{\textstyle{5 \over 2}}}}
\def\quarter{\ifmath{{\textstyle{1 \over 4}}}}
\def\threefifths{\ifmath{{\textstyle{3 \over 5}}}}
\def\teight{\ifmath{{\textstyle{3 \over 8}}}}
\def\tsixteen{\ifmath{{\textstyle{3 \over16}}}}
\def\eighth{\ifmath{{\textstyle{1 \over 8}}}}
\def\ethirds{\ifmath{{\textstyle{8 \over 3}}}}
\def\tnineths{\ifmath{{\textstyle{32\over 9}}}}
\def\fthirds{\ifmath{{\textstyle{4 \over 3}}}}
\def\enineths{\ifmath{{\textstyle{8 \over 9}}}}
\def\threehalf{\ifmath{{\textstyle{3 \over 2}}}}
\def\twothirds{\ifmath{{\textstyle{2 \over 3}}}}
\def\sixth{\ifmath{{\textstyle{1 \over 6}}}}
\def\mlsq{m\ls{L}^2}
\def\mtsq{m\ls{T}^2}
\def\mdsq{m\ls{D}^2}
\def\mssq{m\ls{S}^2}
\def\calmm{{\calm}^2}
\def\cbb{\cos^2\beta}
\def\sbb{\sin^2\beta}
\def\MS{M_{\rm susy}}
\def\MSS{M_{\rm susy}^2}
\def\app#1#2#3{{\sl Act. Phys. Pol. }{\bf B#1} (#2) #3}
\def\apa#1#2#3{{\sl Act. Phys. Austr.}{\bf #1} (#2) #3}
\def\ppnp#1#2#3{{\sl Prog. Part. Nucl. Phys. }{\bf #1} (#2) #3}
\def\npb#1#2#3{{\sl Nucl. Phys. }{\bf B#1} (#2) #3}
\def\jpa#1#2#3{{\sl J. Phys. }{\bf A#1} (#2) #3}
\def\plb#1#2#3{{\sl Phys. Lett. }{\bf B#1} (#2) #3}
\def\prd#1#2#3{{\sl Phys. Rev. }{\bf D#1} (#2) #3}
\def\pR#1#2#3{{\sl Phys. Rev. }{\bf #1} (#2) #3}
\def\prl#1#2#3{{\sl Phys. Rev. Lett. }{\bf #1} (#2) #3}
\def\prc#1#2#3{{\sl Phys. Reports }{\bf #1} (#2) #3}
\def\cpc#1#2#3{{\sl Comp. Phys. Commun. }{\bf #1} (#2) #3}
\def\nim#1#2#3{{\sl Nucl. Inst. Meth. }{\bf #1} (#2) #3}
\def\pr#1#2#3{{\sl Phys. Reports }{\bf #1} (#2) #3}
\def\sovnp#1#2#3{{\sl Sov. J. Nucl. Phys. }{\bf #1} (#2) #3}
\def\jl#1#2#3{{\sl JETP Lett. }{\bf #1} (#2) #3}
\def\jet#1#2#3{{\sl JETP Lett. }{\bf #1} (#2) #3}
\def\zpc#1#2#3{{\sl Z. Phys. }{\bf C#1} (#2) #3}
\def\ptp#1#2#3{{\sl Prog.~Theor.~Phys.~}{\bf #1} (#2) #3}
\def\nca#1#2#3{{\sl Nouvo~Cim.~}{\bf#1A} (#2) #3}
\def\hpa#1#2#3{{\sl Helv.~Phys.~Acta~}{\bf #1} (#2) #3}
\def\aop#1#2#3{{\sl Ann.~of~Phys.~}{\bf #1} (#2) #3}
\def\fP#1#2#3{{\sl Fortschr.~Phys.~}{\bf #1} (#2) #3}
%
\setcounter{page}{1}
\pagestyle{plain}
\hbox to \hsize{}
\vskip1cm
\centerline{\large\bf   Higgs Boson Masses and Couplings}
\vskip5pt
\centerline{\large\bf in the Minimal
Supersymmetric Model}
\vskip1cm
\centerline{Howard E. Haber}
\SCIPP

\vskip1cm
\begin{abstract}
The Higgs sector of the Minimal Supersymmetric Model (MSSM) is a
CP-conserving two-Higgs doublet model that depends, at tree-level,
on two Higgs sector parameters.  In order to accurately determine the
phenomenological implications of this model, one must include
the effects of radiative corrections.
The leading contributions to the one-loop radiative corrections are
exhibited;  large logarithms are resummed by the renormalization group
method.  Implications for Higgs phenomenology are briefly discussed.
\end{abstract}

\section{Introduction}  \label{sec:intro}

The Standard Model with minimal
Higgs content is not expected to be the ultimate
theoretical structure responsible for electroweak symmetry breaking
\cite{hhg,habertasi}.
If the Standard Model is embedded in a more fundamental structure
characterized by a much larger energy scale
(\eg, the Planck scale, which must appear in any theory
of fundamental particles and interactions that includes
gravity), the Higgs boson
would tend to acquire mass of order the largest energy scale due to
radiative corrections.  Only by adjusting (\ie, ``fine-tuning'')
the parameters of the
Higgs potential ``unnaturally'' can one arrange a large hierarchy
between the Planck scale and the scale of electroweak symmetry breaking
\cite{thooft,suss}.
The Standard Model provides no mechanism for this, but supersymmetric
theories have the potential to address these issues.
In a supersymmetric theory, the size of
radiative corrections to scalar squared-masses
is limited by the exact cancelation of quadratically divergent
contributions from loops of particles and their supersymmetric partners.
Since supersymmetry is not
an exact symmetry at low energies, this cancelation must be incomplete,
and the size of the radiative corrections to the Higgs mass is controlled by
the extent of the supersymmetry breaking.  
The resolution of the naturalness
and hierarchy problems requires that the scale of supersymmetry
breaking should not exceed ${\cal O}$(1~TeV) \cite{susysol}.
Such ``low-energy''
supersymmetric theories are especially interesting in that, to date,
they provide the only theoretical framework
in which the problems of naturalness
and hierarchy can be resolved while retaining the Higgs bosons as truly
elementary weakly coupled spin-0 particles.

The Minimal Supersymmetric extension of the Standard Model (MSSM)
contains the Standard Model particle spectrum and the corresponding
supersymmetric partners \cite{susyrev,hehtasi}.
In addition, the MSSM must possess two Higgs doublets in
order to give masses to up and down type fermions in a manner consistent
with supersymmetry (and to avoid gauge anomalies introduced by
the fermionic superpartners of the Higgs bosons).
In particular, the MSSM Higgs sector is a CP-conserving
two-Higgs-doublet model, which can be parametrized at tree-level
in terms of
two Higgs sector parameters. This structure arises
due to constraints imposed by
supersymmetry that determine the Higgs quartic couplings in terms
of electroweak gauge coupling constants.

In section 2, I review the general structure of the (nonsupersymmetric)
two-Higgs-doublet extension of the Standard Model.  By imposing
the constraints of supersymmetry on the quartic terms of the Higgs
potential (and the Higgs-fermion interaction) one obtains the Higgs
sector of the MSSM.  The tree-level predictions of this model are
briefly summarized in section 3.
The inclusion of radiative corrections in the analysis of the
MSSM Higgs sector can have profound implications.
The most dramatic effect of the radiative corrections on
the MSSM Higgs sector is the modification of the tree-level mass
relations of the model.  The leading one-loop
radiative corrections to MSSM Higgs masses
are described in section 4.  These include the full set of one-loop
leading logarithmic terms, and the leading third generation
squark-mixing corrections.  In section 5, the leading logarithms are
resummed to
all orders via the renormalization group technique.  A simple analytic
formula is exhibited which serves as an excellent approximation to the
numerically integrated renormalization group equations.  Numerical
examples demonstrate that the Higgs masses computed in this
approximation lie within 2 GeV of their actual values over a very
large fraction of the supersymmetric parameter space.
Finally, some implications of the radiatively-corrected
Higgs sector are briefly explored in section 6.  Certain technical
details are relegated to the appendices.

\section{The Two-Higgs Doublet Model} \label{sec:two}      

I begin with a brief review of the general (non-supersymmetric)
two-Higgs doublet extension of the Standard Model \cite{hhgref}.
Let $\Phi_1$ and
$\Phi_2$ denote two complex $Y=1$, SU(2)$\ls{L}$ doublet scalar fields.
The most general gauge invariant scalar potential is given by

\vbox{%
\beqno
\calv&=m_{11}^2\Phi_1^\dagger\Phi_1+m_{22}^2\Phi_2^\dagger\Phi_2
-[m_{12}^2\Phi_1^\dagger\Phi_2+{\rm h.c.}]\nonumber\\[6pt]
&\quad +\half\lambda_1(\Phi_1^\dagger\Phi_1)^2
+\half\lambda_2(\Phi_2^\dagger\Phi_2)^2
+\lambda_3(\Phi_1^\dagger\Phi_1)(\Phi_2^\dagger\Phi_2)
+\lambda_4(\Phi_1^\dagger\Phi_2)(\Phi_2^\dagger\Phi_1)
\nonumber\\[6pt]
&\quad +\left\{\half\lambda_5(\Phi_1^\dagger\Phi_2)^2
+\big[\lambda_6(\Phi_1^\dagger\Phi_1)
+\lambda_7(\Phi_2^\dagger\Phi_2)\big]
\Phi_1^\dagger\Phi_2+{\rm h.c.}\right\}\,. \label{pot}
\eeqno }

\noindent
In most discussions of two-Higgs-doublet models, the terms proportional
to $\lambda_6$ and $\lambda_7$ are absent.  This can be achieved by
imposing a discrete symmetry $\Phi_1\rta -\Phi_1$ on the model.  Such a
symmetry would also require $m_{12}=0$ unless we allow a
soft violation of this discrete symmetry by dimension-two terms.\footnote{%
This latter requirement is sufficient to guarantee the absence of
Higgs-mediated tree-level flavor changing neutral currents.}
For the moment, I will refrain from setting any of the coefficients
in eq.~(\ref{pot}) to zero.  In principle, $m_{12}^2$, $\lambda_5$,
$\lambda_6$ and $\lambda_7$ can be complex.  However, for simplicity, I
shall ignore the possibility of CP-violating effects in the Higgs sector
by choosing all coefficients in eq.~(\ref{pot}) to be real.
The scalar fields will
develop non-zero vacuum expectation values if the mass matrix
$m_{ij}^2$ has at least one negative eigenvalue. Imposing CP invariance
and U(1)$\ls{\rm EM}$ gauge symmetry, the minimum of the potential is
\beq
\langle \Phi_1 \rangle={1\over\sqrt{2}} \left(
\begin{array}{c} 0\\ v_1\end{array}\right), \qquad \langle
\Phi_2\rangle=
{1\over\sqrt{2}}\left(\begin{array}{c}0\\ v_2
\end{array}\right)\,,\label{potmin}
\eeq
where the $v_i$ are assumed to be real.  It is convenient to introduce
the following notation:
\beq
v^2\equiv v_1^2+v_2^2={4\mw^2\over g^2}=(246~{\rm GeV})^2\,,
\qquad\qquad\tb\equiv\tanb\equiv{v_2\over v_1}\,.\label{tanbdef}
\eeq

Of the original eight scalar degrees of freedom, three Goldstone
bosons ($G^\pm$ and $G^0$)
are absorbed (``eaten'') by the $W^\pm$ and $Z$.  The remaining
five physical Higgs particles are: two CP-even scalars ($\hl$ and
$\hh$, with $\mhl\leq \mhh$), one CP-odd scalar ($\ha$) and a charged
Higgs pair ($\hpm$). The mass parameters $m_{11}$ and $m_{22}$ can be
eliminated by minimizing the scalar potential.  The resulting
squared masses for the CP-odd and charged Higgs states are
\beqno
\mha^2 &={m_{12}^2\over \sb\cb}-\half
v^2\big(2\lambda_5+\lambda_6\tb^{-1}+\lambda_7\tb\big)\,,\nonumber\\[6pt]
m_{H^{\pm}}^2 &=m_{A^0}^2+\half v^2(\lambda_5-\lambda_4)\,.
\label{mamthree}
\eeqno
\vskip1pc

\noindent
The two CP-even Higgs states mix according to the following squared mass
matrix:
\vbox{%
\beqno
\calm^2 &=m_{A^0}^2  \left(
   \begin{array}{cc}  \sb^2& -\sb\cb\\
                     -\sb\cb& \cb^2 \end{array}\right) \nonumber\\[3pt]
&+v^2 \left( \begin{array}{cc}
  \lambda_1\cb^2+2\lambda_6\sb\cb+\lambda_5\sb^2
    &(\lambda_3+\lambda_4)\sb\cb+\lambda_6 \cb^2+\lambda_7\sb^2 \\[3pt]
 (\lambda_3+\lambda_4)\sb\cb+\lambda_6  \cb^2+\lambda_7\sb^2
    &\lambda_2\sb^2+2\lambda_7\sb\cb+\lambda_5\cb^2
\end{array}\right) \,,
\label{massmhh}
\eeqno }

\noindent
where $s_\beta\equiv\sin\beta$ and $c_\beta\equiv\cos\beta$.
The physical mass eigenstates are
\beqno
\hh &=(\sqrt{2}{\rm Re\,}\Phi_1^0-v_1)\cos\alpha+
(\sqrt{2}{\rm Re\,}\Phi_2^0-v_2)\sin\alpha\,,\nonumber\\
\hl &=-(\sqrt{2}{\rm Re\,}\Phi_1^0-v_1)\sin\alpha+
(\sqrt{2}{\rm Re\,}\Phi_2^0-v_2)\cos\alpha\,.
\label{scalareigenstates}
\eeqno
The corresponding masses are
\beq
 m^2_{\hh,\hl}=\half\left[{\cal M}_{11}^2+{\cal M}_{22}^2
\pm \sqrt{({\cal M}_{11}^2-{\cal M}_{22}^2)^2 +4({\cal M}_{12}^2)^2}
\ \right]\,,
\label{higgsmasses}
\eeq
and the mixing angle $\alpha$ is obtained from
\beqno
\sin 2\alpha &={2{\cal M}_{12}^2\over
\sqrt{({\cal M}_{11}^2-{\cal M}_{22}^2)^2 +4({\cal M}_{12}^2)^2}}\ ,\nonumber\\
\cos 2\alpha &={{\cal M}_{11}^2-{\cal M}_{22}^2\over
\sqrt{({\cal M}_{11}^2-{\cal M}_{22}^2)^2 +4({\cal M}_{12}^2)^2}}\ .
\label{alphadef}
\eeqno

The phenomenology of the two-Higgs doublet model depends in detail on
the various couplings of the Higgs bosons to gauge bosons, Higgs
bosons and fermions.  The Higgs couplings to gauge bosons follow from
gauge invariance and are thus model independent.  For example,
the couplings of the two CP-even Higgs bosons to $W$ and $Z$ pairs
are given in terms of the angles $\alpha$ and $\beta$ by
\beqno
 g\ls{\hl VV}&=g\ls{V} m\ls{V}\sinbma \nonumber \\[3pt]
           g\ls{\hh VV}&=g\ls{V} m\ls{V}\cosbma\,,\label{vvcoup}
\eeqno
where
\beq
g\ls V\equiv\begin{cases}
g,& $V=W\,$,\\ g/\cos\theta_W,& $V=Z\,$. \end{cases}
\label{hix}
\eeq
There are no tree-level couplings of $\ha$ or $\hpm$ to $VV$.
Gauge invariance also determines the strength of the trilinear
couplings of one gauge boson to two Higgs bosons.  For example,
\beqno g\ls{\hl\ha Z}&={g\cosbma\over 2\cos\theta_W}\,,\nonumber \\[3pt]
           g\ls{\hh\ha Z}&={-g\sinbma\over 2\cos\theta_W}\,.
           \label{hvcoup}
\eeqno

In the examples shown above, some of the couplings can be suppressed
if either $\sin(\beta-\alpha)$ or $\cos(\beta-\alpha)$ is very small.
Note that all the vector boson--Higgs boson couplings
cannot vanish simultaneously. From the expressions above, we see
that the following sum rules must hold separately for $V=W$ and $Z$:
\beqno
g_{\hh V V}^2 + g_{\hl V V}^2 &=
                      g\ls{V}^2m\ls{V}^2\,,\nonumber \\[3pt]
g_{\hl\ha Z}^2+g_{\hh\ha Z}^2&=
             {g^2\over 4\cos^2\theta_W}\,.
\label{sumruletwo}
\eeqno
These results are a
consequence of the tree-unitarity of the electroweak theory~%
\cite{ghw}.
Moreover, if we focus on a given CP-even Higgs state, we note that
its couplings to $VV$ and $\ha V$ cannot be simultaneously
suppressed, since eqs.~(\ref{vvcoup})--(\ref{hvcoup}) imply that
\beq
  g^2_{h ZZ} + 4m^2_Z g^2_{hA^0Z} = {g^2m^2_Z\over
        \cos^2\theta_W}\,,\label{hxi}
\eeq
for $h=\hl$ or $\hh$.  Similar considerations also hold for
the coupling of $\hl$ and $\hh$ to $W^\pm H^\mp$.  We can summarize
the above results by noting that the coupling of $\hl$ and $\hh$ to
vector boson pairs or vector--scalar boson final states is proportional
to either $\sin(\beta-\alpha)$ or $\cos(\beta-\alpha)$ as indicated
below \cite{hhg,hhgsusy}.
\beq
\renewcommand{\arraycolsep}{2cm}
\let\us=\underline
\begin{array}{ll}
  \us{\cos(\beta-\alpha)}&  \us{\sin(\beta-\alpha)}\\[3pt]
       \hh W^+W^-&        \hl W^+W^- \\
       \hh ZZ&            \hl ZZ \\
       Z\ha\hl&          Z\ha\hh \\
       W^\pm H^\mp\hl&  W^\pm H^\mp\hh \\
       ZW^\pm H^\mp\hl&  ZW^\pm H^\mp\hh \\
       \gamma W^\pm H^\mp\hl&  \gamma W^\pm H^\mp\hh
\end{array}
\label{littletable}
\eeq
Note in particular that {\it all} vertices
in the theory that contain at least
one vector boson and {\it exactly one} non-minimal Higgs boson state
($\hh$, $\ha$ or $\hpm$) are proportional to $\cos(\beta-\alpha)$.

The 3-point and 4-point Higgs self-couplings depend on the
parameters of the
two-Higgs-doublet potential [eq.~(\ref{pot})].  The Feynman rules for
the trilinear Higgs vertices are listed in Appendix A.
The Feynman rules for the 4-point Higgs vertices are rather
tedious in the general two-Higgs-doublet
model and will not be given here.

The Higgs couplings
to fermions are model dependent, although their form is often
constrained by discrete symmetries that are imposed in order to
avoid tree-level flavor changing neutral currents mediated by Higgs
exchange \cite{gw}.
An example of a model that respects this constraint is one in
which one Higgs doublet (before symmetry
breaking) couples exclusively to down-type fermions and the other
Higgs doublet couples exclusively to up-type
fermions.  This is the pattern of couplings found in the MSSM.  The
results in this case are as follows.  The couplings
of the neutral Higgs bosons to $f\bar f$
relative to the Standard Model
value, $gm_f/2\mw$, are given by (using 3rd family notation)
\begin{eqaligntwo} \label{qqcouplings}
\hl b\bar b:&~~~ -{\sin\alpha\over\cos\beta}=\sin(\beta-\alpha)
-\tan\beta\cos(\beta-\alpha)\,,\nonumber\\[3pt]
\hl t\bar t:&~~~ \phm{\cos\alpha\over\sin\beta}=\sin(\beta-\alpha)
+\cot\beta\cos(\beta-\alpha)\,,\nonumber\\[3pt]
\hh b\bar b:&~~~ \phm{\cos\alpha\over\cos\beta}=\cos(\beta-\alpha)
+\tan\beta\sin(\beta-\alpha)\,,\nonumber\\[3pt]
\hh t\bar t:&~~~ \phm{\sin\alpha\over\sin\beta}=\cos(\beta-\alpha)
-\cot\beta\sin(\beta-\alpha)\,,\nonumber\\[3pt]
\ha b \bar b:&~~~\phm\gamma_5\,{\tan\beta}\,,\nonumber\\[3pt]
\ha t \bar t:&~~~\phm\gamma_5\,{\cot\beta}\,,
\end{eqaligntwo}
(the $\gamma_5$ indicates a pseudoscalar coupling), and the
charged Higgs boson coupling to fermion pairs
(with all particles pointing into the vertex) is given by
\beq
g_{H^- t\bar b}={g\over{2\sqrt{2}\mw}}\
[m_t\cot\beta\,(1+\gamma_5)+m_b\tan\beta\,(1-\gamma_5)].
\label{hpmqq}
\eeq

The pattern of couplings displayed above can be understood in the
context of the {\it decoupling limit} of the two-Higgs-doublet model
\cite{habernir,DECP}.
First, consider the Standard Model Higgs boson ($\hsm$).  At tree-level,
the
Higgs self-coupling is related to its mass.  If $\lambda$ is the quartic
Higgs self-interaction strength [see eq.~(\ref{potsm})], then $\lambda=
\mhsm^2/v^2$.
This means that one cannot take $\mhsm$ arbitrarily large without the
attendant growth in $\lambda$.  That is, the heavy Higgs limit in the
Standard Model exhibits non-decoupling.  In models of a non-minimal
Higgs sector, the situation is more complex. In some models
(with the Standard Model as one example), it is not
possible to take any Higgs mass much larger than ${\cal O}(v)$ without
finding at least one strong Higgs self-coupling.
In other models, one
finds that the non-minimal Higgs boson masses can be taken large at
fixed Higgs self-couplings.
Such behavior can arise in models that possess one (or more)
off-diagonal squared-mass parameters in addition to the diagonal scalar
squared-masses. In the limit where the off-diagonal squared-mass
parameters are taken large [keeping the dimensionless
Higgs self-couplings fixed and $\lsim{\cal O}(1)$], the heavy Higgs
states decouple, while both light and heavy Higgs bosons remain
weakly-coupled.  In this decoupling limit,
exactly one neutral CP-even Higgs
scalar remains light, and its properties are precisely those of the
(weakly-coupled) Standard Model Higgs boson.  That is,
$\hl\simeq\hsm$, with
$\mhl\sim{\cal O}(\mz)$, and all other non-minimal Higgs states are
significantly heavier than $\mhl$.  Squared-mass splittings of the heavy
Higgs
states are of ${\cal O}(\mz^2)$, which means that all heavy Higgs states
are approximately degenerate, with mass differences of order
$\mz^2/\mha$ (here $\mha$ is approximately equal to the common heavy
Higgs mass scale).  In contrast, if the non-minimal Higgs sector is
weakly coupled but far from the decoupling limit, then $\hl$ is not
separated in mass from the other Higgs states.  In this case, the
properties\footnote{The basic property of the Higgs coupling strength
proportional to mass is maintained.  But,
the precise coupling strength patterns of
$\hl$ will differ from those of $\hsm$ in the non-decoupling limit.}
of $\hl$ differ significantly from those of $\hsm$.

Below, I exhibit the decoupling limit of the most general CP-even
two-Higgs-doublet model \cite{DECP}.
It is convenient to define four squared mass combinations:
\beqno
\mlsq\equiv&\ \calm^2_{11}\cos^2\beta+\calm^2_{22}\sin^2\beta
+\calm^2_{12}\sin2\beta\,,\nonumber \\[3pt]
\mdsq\equiv&\ \left(\calm^2_{11}\calm^2_{22}-\calm^4_{12}\right)^{1/2}\,,
\nonumber \\[3pt]
\mtsq\equiv&\ \calm^2_{11}+\calm^2_{22}\,,\nonumber \\[3pt]
\mssq\equiv&\ \mha^2+\mtsq\,,
\label{massdefs}
\eeqno
in terms of the elements of the neutral CP-even Higgs squared-mass
matrix [eq.~(\ref{massmhh})].  In terms of the above quantities,
\beq
 m^2_{\hh,\hl}=\half\left[\mssq\pm\sqrt{m\ls{S}^4-4\mha^2\mlsq
-4m\ls{D}^4}\,\right]\,,
\label{cpevenhiggsmasses}
\eeq
and
\beq
\cos^2(\beta-\alpha)= {\mlsq-\mhl^2\over\mhh^2-\mhl^2}\,.
\label{cosbmasq}
\eeq
In the decoupling limit,
all the Higgs self-coupling constants $\lambda_i$ are
held fixed such that $\lambda_i\lsim1$, while taking
$\mha^2\gg\lambda_iv^2$.
Then $\calm^2_{ij}\sim{\cal O}(v^2)$, and
it follows that:
\beq
\mhl\simeq m\ls{L}\,,\qquad\qquad \mhh\simeq\mha\simeq\mhpm\,,
\label{approxmasses}
\eeq
and
\beq
\cos^2(\beta-\alpha)\simeq\, {\mlsq(\mtsq-\mlsq)-m\ls{D}^4\over\mha^4}\,.
\label{approxcosbmasq}
\eeq
\vskip9pt\noindent
Note that eq.~(\ref{approxcosbmasq}) implies that
$\cos(\beta-\alpha)=
{\cal O}(\mz^2/\mha^2)$ in the decoupling limit, which means that
the $\hl$ couplings to
Standard Model particles match precisely those of the Standard Model
Higgs boson.  These results are easily confirmed by
considering the $\cos(\beta-\alpha)\to 0$ limit of
eqs.~(\ref{vvcoup})--(\ref{qqcouplings}).

Although no experimental evidence for the Higgs boson yet exists, there
are some experimental as well as theoretical constraints on the
parameters of the
two-Higgs doublet model.  Experimental limits on the charged and
neutral Higgs masses have been obtained at LEP.
For the charged Higgs boson, $\mhpm>44$~GeV \cite{LEPHIGGS}.
This is the most model-independent bound and assumes only that
the $\hpm$ decays dominantly into $\tau^+\nu_\tau$, $c \bar s$
and $c\bar b$.
The LEP limits on the masses of $\hl$ and $\ha$ are obtained by searching
simultaneously for $e^+e^- \to h^0 f\bar f$ and $e^+e^- \to h^0 A^0$,
which are mediated by $s$-channel $Z$-exchange \cite{janot}.
The $ZZ\hl$ and $Z\hl\ha$
couplings that govern these two decay rates are proportional to
$\sin(\beta-\alpha)$ and $\cos(\beta-\alpha)$, respectively.
Thus, one can use the
LEP data to deduce limits on $\mhl$ and $\mha$ as a function of
$\sin(\beta-\alpha)$.
Stronger limits can be obtained in the MSSM where
$\sin(\beta-\alpha)$ is determined by other model parameters.
At present, taking into account data from LEP-1 and
the most recent LEP-2 data (at $\sqrt{s}=161$ and 172~GeV),
one can exclude the MSSM Higgs mass
ranges: $\mhl< 62.5$~GeV (independent of the value of $\tan\beta$)
and $\mha< 62.5$~GeV (assuming $\tan\beta> 1$)
\cite{ypan}.

The experimental information on the
parameter $\tanb$ is quite meager.
For definiteness, let us assume that the Higgs-fermion couplings are
specified as in eq.~(\ref{qqcouplings}).
The Higgs coupling to top quarks
is proportional to $gm_t/2\mw$, and is
therefore the strongest of all Higgs-fermion couplings.
For $\tanb<1$, the Higgs couplings to top-quarks
are further enhanced by a factor of $1/\tanb$.  As a result, some
experimental limits on $\tanb$ exist based on the
non-observation of virtual effects involving the $H^-t\bar b$
coupling.  Clearly, such limits depend both on $\mhpm$ and $\tanb$.
The most sensitive limits are obtained from the measurements of
$B^0$-$\overline{B^0}$ mixing and the widths of $b\to s\gamma$ and $Z\to
b\bar b$ \cite{grant}.  For example,
the process $b\to s\gamma$ can be significantly enhanced
due to charged Higgs boson exchange.  If there are no other competing
non-Standard Model contributions (and this is a big {\it if}), then
present data excludes charged Higgs masses less than about
250 GeV \cite{joanne} (independent of the value of $\tan\beta$).  In
some regions of $\tan\beta$, the limits on the charged Higgs mass can be
even more severe.  However,
other virtual contributions may
exist that can cancel the effects of the charged Higgs exchange.
For example, in the MSSM, constraints on $\tanb$ and $\mhpm$ are
significantly
weaker.  For $\tanb\gg 1$, the Higgs couplings to bottom-quarks are
enhanced by a factor of $\tanb$.  In this case, the measured rate
for the inclusive
decay of $B\to X+\tau\nu_\tau$ can be used to set an upper limit on
$\tanb$ as a function of the charged Higgs mass.  This is accomplished
by setting a limit on the contribution of the {\it tree-level} charged
Higgs exchange.  Present data can be used to set a $2\sigma$ upper bound
of $\tanb< 42(\mhpm/\mw)$ \cite{ghn}.  In the MSSM, this bound could
be weakened due to one-loop QCD corrections mediated by the exchange of
supersymmetric particles \cite{sola}.

Theoretical considerations also lead to bounds on $\tanb$.  The crudest
bounds arise from unitarity constraints.  If $\tanb$ becomes too small,
then the Higgs coupling to top quarks becomes too strong.
In this case, the
tree-unitarity of processes involving the Higgs-top quark Yukawa
coupling is violated.  Perhaps this should not be
regarded as a theoretical defect, although it does render any
perturbative analysis unreliable.  A rough lower bound advocated
by Ref.~\cite{hewett}, $\tanb\gsim 0.3$, corresponds to a
Higgs-top quark coupling in the perturbative region.  A similar argument
involving the Higgs-bottom quark coupling would yield $\tanb\lsim 120$.
A more solid theoretical constraint is based on the
requirement that Higgs--fermion Yukawa
couplings remain finite when running
from the electroweak scale to some large energy scale $\Lambda$.
Above $\Lambda$, one assumes that new physics enters.  The
limits on $\tanb$ depend on $\mt$ and
the choice of the high energy scale $\Lambda$.
Using the renormalization group equations given in Appendix B,
one integrates from the electroweak scale to $\Lambda$ (allowing for
the possible existence of a supersymmetry-breaking
scale, $\mz\leq\msusy\leq
\Lambda$), and determines the region of $\tanb$--$\mt$ parameter space
in which the Higgs-fermion Yukawa couplings remain finite.
This exercise has recently been carried out at two-loops in
Ref.~\cite {schrempp}.  Suppose that the low-energy theory at
the electroweak scale is the MSSM, and that
there is no additional new physics below the grand unification scale of
$\Lambda=2\times 10^{16}$~GeV.  Then, for $\mt=170$~GeV,
the Higgs-fermion Yukawa couplings remain finite at all energy scales below
$\Lambda$ if $1.5\lsim\tanb\lsim
65$. Note that this result is consistent with the scenario of radiative
electroweak
symmetry breaking in low-energy supersymmetry based on supergravity,
which requires that $1\lsim\tanb\lsim \mt/\mb$.

\section{The Higgs Sector of the MSSM at Tree Level} \label{sec:three}

The Higgs sector of the MSSM is a CP-conserving
two-Higgs-doublet model, with a Higgs potential whose dimension-four
terms respect supersymmetry and with
restricted Higgs-fermion couplings in which $\Phi_1$ couples exclusively
to down-type fermions while $\Phi_2$ couples exclusively to up-type
fermions \cite{hhgref}.  Using the notation of eq.~(\ref{pot}), the quartic
couplings $\lambda_i$ are given by
\beqno%
\lambda_1 &=\lambda_2 = \fourth (g^2+g'^2)\,, \nonumber\\
\lambda_3 &=\fourth (g^2-g'^2)\,,    \nonumber\\
\lambda_4 &=-\half g^2\,, \nonumber\\
\lambda_5 &=\lambda_6=\lambda_7=0\,.\label{bndfr}
\eeqno
Inserting these results into eqs.~(\ref{mamthree}) and (\ref{massmhh}), it
follows that

\vbox{%
\beqno%
\mha^2 &=m_{12}^2(\tan\beta+\cot\beta)\,,\nonumber\\
\mhpm^2 &=\mha^2+\mw^2\,,
\label{susymhpm}
\eeqno
}

\noindent and the tree-level neutral CP-even mass matrix is given by
\beq
{\cal M}_0^2 =    \left(
\begin{array}{ll}
\mha^2 \sin^2\beta + m^2_Z \cos^2\beta&
          -(\mha^2+m^2_Z)\sin\beta\cos\beta\\
 -(\mha^2+m^2_Z)\sin\beta\cos\beta&
 \mha^2\cos^2\beta+ m^2_Z \sin^2\beta\end{array}\right)\,.\label{kv}
\eeq
The eigenvalues of ${\cal M}_0^2$ are
the squared masses of the two CP-even Higgs scalars
\beq
  m^2_{H^0,h^0} = \half \left( \mha^2 + m^2_Z \pm
                  \sqrt{(\mha^2+m^2_Z)^2 - 4m^2_Z \mha^2 \cos^2 2\beta}
                  \; \right)\,.\label{kviii}
\eeq
and the diagonalizing angle is $\alpha$, with
\beq
  \cos 2\alpha = -\cos 2\beta \left( {\mha^2-m^2_Z \over
                  m^2_{H^0}-m^2_{h^0}}\right)\,,\qquad
  \sin 2\alpha = -\sin 2\beta \left( m^2_{H^0} + m^2_{h^0} \over
                   m^2_{H^0}-m^2_{h^0} \right)\,.\label{kix}
\eeq
From these results,it is easy to obtain:
\beq
\cos^2(\beta-\alpha)={\mhl^2(\mz^2-\mhl^2)\over
\mha^2(\mhh^2-\mhl^2)}\,.
\label{cbmasq}
\eeq
Thus, in the MSSM, two parameters (conveniently chosen to be $\mha$
and $\tanb$) suffice to fix all other tree-level Higgs sector parameters.

Consider the decoupling limit where $\mha\gg\mz$.  Then, the above
formulae yield
\beqno
\mhl^2\simeq\ &\mz^2\cos^2 2\beta\,,\nonumber \\[3pt]
\mhh^2\simeq\ &\mha^2+\mz^2\sin^2 2\beta\,,\nonumber \\[3pt]
\mhpm^2=\ & \mha^2+\mw^2\,,\nonumber \\[3pt]
\cos^2(\beta-\alpha)\simeq\ &{\mz^4\sin^2 4\beta\over 4\mha^4}\,.
\label{largema}
\eeqno
Two consequences are immediately apparent.
First, $\mha\simeq\mhh
\simeq\mhpm$, up to corrections of ${\cal O}(\mz^2/\mha)$.  Second,
$\cos(\beta-\alpha)=0$ up to corrections of ${\cal O}(\mz^2/\mha^2)$.
Of course, these results were expected based on the discussion of the
decoupling limit in the general two-Higgs-doublet model given in
section 2.

Finally, a number of important mass inequalities can be derived
from the expressions for the tree-level Higgs masses obtained above,

\vbox{%
\beqno
  m_{h^0} &\leq \mha \nonumber \\
  m_{h^0} &\leq m|\cos 2\beta | \leq m_Z \,,
            \qquad {\rm with}\ m \equiv min(m_Z,\mha) \nonumber\\
  m_{H^0} &\geq m_Z\,, \nonumber\\
m_{H^\pm} &\geq\mw\,. \label{kx}
\eeqno
}

\section{Radiative Corrections to the MSSM Higgs Masses}     

\subsection{Overview}

The tree-level results of the previous section are modified when
radiative corrections are incorporated.  Naively, one might expect
radiative corrections to have a minor effect on the phenomenological
implications of the model.  However, in the MSSM, some of the
tree-level Higgs mass relations may be significantly changed at
one-loop, with profound implications for the phenomenology.  For
example, consider the tree-level bound on the lightest CP-even Higgs boson
of the MSSM: $\mhl\leq\mz|\cos 2\beta|\leq\mz$.
The LEP-2 collider (running at its projected maximum
center-of-mass energy of 192~GeV, with an integrated luminosity of
150~${\rm pb}^{-1}$) will discover at least one Higgs boson of the MSSM
if $\mhl\leq\mz$ \cite{janot}.  Thus, if the tree-level Higgs mass bound
holds, then the absence of a Higgs discovery at LEP would
rule out the MSSM.  However, when radiative corrections are included,
the light Higgs mass upper bound may be increased significantly.  In the
one-loop leading logarithmic approximation \cite{hhprl,early-veff}
\begin{equation} \label{mhlapprox}
\mhl^2\lsim\mzz\cos^2\beta+{3g^2 m_t^4\over
8\pi^2\mw^2}\,\ln\left({M_{\tilde t_1}M_{\tilde t_2}\over
\mt^2}\right)\,,
\end{equation}
where $M_{\tilde t_1}$, $M_{\tilde t_2}$ are the masses of the two
top-squark mass eigenstates. Observe that the Higgs mass upper bound
is very sensitive to the top mass and depends logarithmically on the
top-squark masses.  In addition, due to the increased upper bound for
$\mhl$, the non-observation
of a Higgs boson at LEP-2 cannot rule out the MSSM.

Although eq.~(\ref{mhlapprox}) provides a rough
guide to the Higgs mass upper bound, it is not sufficiently precise for
LEP-2 phenomenology, whose Higgs mass reach depends delicately on the
MSSM parameters.  In addition, in order to perform precision Higgs
measurements and make comparisons with theory, more accurate results
for the Higgs sector masses (and couplings) are required.
The radiative corrections to the Higgs mass have been computed by a
number of techniques, and using a variety of approximations such as
effective potential \cite{early-veff,veff,berz,erz,carena} and
diagrammatic methods
\cite{hhprl,turski,brig,madiaz,1-loop,hempfhoang,completeoneloop}.
Complete one-loop diagrammatic computations of the MSSM Higgs masses
have been presented by a number of groups \cite{completeoneloop};
the resulting expressions are quite complex,
and depend on all the parameters of the MSSM.
(The dominant two-loop next-to-leading
logarithmic results are also known~\cite{hempfhoang}.)
Moreover, as noted above, the largest contribution to the one-loop radiative
corrections is enhanced by a factor of $m_t^4$ and grows
logarithmically with the top squark mass.  Thus, higher order
radiative corrections can be non-negligible for large top
squark masses, in which case the large logarithms must be resummed.

The renormalization group (RG) techniques for resumming the leading
logarithms has been developed by a number of
authors \cite{rge,2loopquiros,llog}.
The computation of the RG-improved
one-loop corrections requires numerical integration of a coupled set of
RG equations~\cite{llog}.
Although this program has been
carried out in the literature, the procedure is unwieldy
and not easily amenable to large-scale Monte-Carlo analyses.
Recently, two groups have presented a simple analytic procedure for
accurately approximating $m_{h^0}$.
These methods can be easily implemented, and incorporate both the
leading one-loop and two-loop effects and the RG-improvement.
Also included are the leading effects at one loop of the supersymmetric
thresholds (the most important effects of this type are squark mixing
effects in the third generation). Details of the techniques can
be found in Refs.~\cite{hhh} and \cite{carena}.
Here, I simply quote two specific bounds, assuming
$\mt=175$~GeV and $M_{\tilde t}\lsim 1$~TeV:
$\mhl\lsim 112$~GeV if top-squark mixing is negligible, while
$\mhl\lsim 125$~GeV if top-squark mixing is ``maximal''.
%
%
Maximal mixing corresponds to
an off-diagonal squark squared-mass that produces the largest value of
$\mhl$.  This mixing leads to an extremely large splitting of top-squark
mass eigenstates.

The charged Higgs mass is also constrained in the MSSM.  At tree level,
$\mhpm$ is given by eq.~(\ref{susymhpm}),
which implies that charged Higgs bosons cannot
be pair produced at LEP-2.  Radiative corrections modify the tree-level
prediction, but the corrections are typically smaller than the neutral
Higgs mass corrections discussed above.  Although $\mhpm\geq\mw$ is not
a strict bound when one-loop corrections are included, the bound holds
approximately over most of MSSM parameter space (and can be significantly
violated only when $\tanb$ is well below 1, a region of parameter space
that is theoretically disfavored).

In the remainder of this section, I shall
present formulae which exhibit the leading contributions to
the one-loop corrected Higgs masses.
Symbolically,
\begin{eqnarray}
\mhpm^2 & = & \left(\mhpm^2\right)_{\rm 1LL}
+\left(\Delta \mhpm^2\right)_{\rm mix}\,,\nonumber \\
\calm^2 & = & \calm^2_{\rm 1LL}+ \Delta\calm^2_{\rm mix}\,,
\label{oneloopmasses}
\end{eqnarray}
where the subscript {\sl 1LL} refers to the tree-level plus the
one-loop leading logarithmic approximation to the full one-loop
calculation,
and the subscript {\sl mix} refers to the contributions
arising from $\widetilde q_L$--$\widetilde q_R$ mixing effects of
the third generation squarks.
The CP-even Higgs mass-squared eigenvalues are then obtained by using
eq.~(\ref{higgsmasses})
and the corresponding mixing angle, $\alpha$, is
obtained from eq.~(\ref{alphadef}).

In the simplest approximation, squark mixing effects are
neglected and  the supersymmetric spectrum is characterized by one
scale, called $\msusy$.  We assume that $\msusy$ is sufficiently
large compared to $\mz$ such that logarithmically enhanced terms
at one-loop dominate over the non-logarithmic terms.\footnote{If this
condition does not hold, then the radiative corrections would
constitute only a minor perturbation on the tree-level predictions.}
In this case, the full
one-loop corrections ({\it e.g.}, obtained by a diagrammatic
computation)
are well approximated by the one-loop leading logarithmic approximation.
Next, we incorporate the effects of squark mixing, which constitute the
largest potential source of non-logarithmic one-loop corrections.
In particular, these contributions to the Higgs mass radiative
corrections arise from the exchange of the third generation squarks.
Now, the approximation is parameterized by four supersymmetric
parameters:
$\msusy$ (a common supersymmetric particle mass) and the third
generation squark mixing parameters: $A_t$, $A_b$ and $\mu$.
A more comprehensive set of formulae can be derived by treating the
third generation squark sector more precisely by
accounting for non-degenerate top and bottom squark masses.
This approximation is characterized by seven supersymmetric
parameters---the three squark mixing parameters mentioned above,
three soft-supersymmetry-breaking diagonal squark mass parameters,
$M_Q$, $M_U$, and $M_D$, and a common supersymmetry mass parameter
$\msusy$ which characterizes the masses of the first two generations
of squarks, the sleptons, the charginos, and the neutralinos.

Given an approximation to the one-loop Higgs mass (at some level of
approximation as described above), one must incorporate the
RG-improvement if $\msusy\gg\mz$.  A simple analytic procedure of
Ref.~\cite{hhh} is described in the section 5, and some numerical
results are presented there.  Similar results have also been obtained by
Carena and collaborators, where analytic approximations to the
RG-improved radiatively
corrected MSSM Higgs masses are also developed \cite{carena}.
Although the approaches are somewhat different, the numerical
results (in cases which have been compared) typically agree
to within 1~GeV in the evaluation of Higgs masses.

\subsection{One-Loop Leading Logarithmic Corrections to the MSSM Higgs
Masses}

The leading logarithmic expressions for Higgs masses can be computed
from the one-loop renormalization group equations (RGEs)
of the gauge and Higgs self-couplings, following
Ref.~\cite{llog}.
The method employs eqs.~(\ref{mamthree}) and (\ref{massmhh}),
which are evaluated by treating the $\lambda_i$
as running parameters evaluated at the electroweak scale, $\mweak$.
In addition, we identify the $W$ and $Z$ masses by

\vbox{%
\beqno\mw^2&=\quarter g^2(v_1^2+v_2^2)\,,\nonumber\\
\mz^2&=\quarter (g^2+g'^2)(v_1^2+v_2^2)\,,\label{vmasses}
\eeqno }

\noindent where the running gauge couplings are also evaluated at $\mweak$.
Of course, the gauge couplings, $g$ and $g'$ are known from
experimental measurements which are performed at the scale $\mweak$.
The $\lambda_i(\mweak^2)$ are determined from supersymmetric boundary
conditions at $\msusy$ and RGE running down to $\mweak$. That is, if
supersymmetry were unbroken, then the
$\lambda_i$ would
be fixed according to eq.~(\ref{bndfr}).  Since supersymmetry is broken,
we regard eq.~(\ref{bndfr}) as boundary conditions for the running
parameters, valid at (and above) the energy scale $\msusy$.  That is,
we take

\vbox{%
\beqno
\lambda_1(\msusy^2)&=\lambda_2(\msusy^2)=\fourth[g^2(\msusy^2)
+g'^2(\msusy^2)],\nonumber \\[6pt]
\lambda_3(\msusy^2)&=\fourth\left[g^2(\msusy^2)-g'^2(\msusy^2)\right],
\nonumber \\  [6pt]
\lambda_4(\msusy^2)&=-\half g^2(\msusy^2),\nonumber \\[6pt]
\lambda_5(\msusy^2)&=\lambda_6(\msusy^2)=
\lambda_7(\msusy^2)=0\,,
\label{boundary}
\eeqno
}

\noindent in accordance with the tree-level relations of the MSSM.
At scales below $\msusy$, the gauge and
quartic couplings evolve according to the
renormalization group equations (RGEs) of the non-supersymmetric
two-Higgs-doublet model given in eqs.~(B.5)--(B.7).
These equations
are of the form:
\beq
{dp_i\over dt} =
\beta_i(p_1,p_2,\ldots)\qquad\mbox{with}~t\equiv\ln\,\mu^2
\,,\label{rgeqs}
\eeq
where $\mu$ is the energy scale, and
the $p_i$ are the
parameters of the theory ($p_i = g_j^2,\lambda_k,\ldots$).
The relevant $\beta$-functions can be found in Appendix B.
The boundary conditions together with the RGEs imply that, at the
leading-log level, $\lambda_5$, $\lambda_6$ and $\lambda_7$ are zero
at all energy scales.  Solving the RGEs with
the supersymmetric boundary conditions at $\msusy$, one can determine
the $\lambda_i$ at the weak scale.
The resulting values for $\lambda_i(\mweak)$
are then inserted into eqs.~(\ref{mamthree}) and (\ref{massmhh}) to obtain the
radiatively corrected Higgs masses.  Having solved the one-loop RGEs,
the Higgs masses thus obtained  include the leading
logarithmic radiative corrections summed to all orders in perturbation
theory.

The RGEs can be solved by numerical analysis on the computer.
In order to derive the one-loop leading logarithmic corrections, it is
sufficient to solve the RGEs iteratively.  In first
approximation, we can take the right hand side of eq.~(\ref{rgeqs}) to
be independent of $\mu^2$.  That is, we compute the $\beta_i$ by
evaluating the parameters $p_i$ at the scale $\mu=\msusy$.
Then, integration of the RGEs is
trivial, and we obtain
\beq
p_i(\mweak^2)=p_i(\msusy^2)-\beta_i\,\ln\left({\msusy^2\over\mweak^2}
\right)\,.\label{oneloopllog}
\eeq
This result demonstrates that the first iteration corresponds
to computing
the one-loop radiative corrections in which only terms proportional to
$\ln\msusy^2$ are kept.  It is straightforward to work out the
one-loop leading logarithmic expressions for the $\lambda_i$ and
the Higgs masses.

First consider the charged Higgs mass.  Since
$\lambda_5(\mu^2)=0$ at all scales, we need only consider $\lambda_4$.
Evaluating $\beta_{\lambda_4}$ at $\mu=\msusy$, we compute

\vbox{%
\beqno%
\lambda_4(\mw^2)=-\half g^2
-&{1\over{32\pi^2}}\biggl[\bigl({\textstyle{
4\over 3}}N_g+{\textstyle{1\over 6}}N_H-{\textstyle{10\over 3}}
\bigr)g^4+5g^2g'^2  \nonumber\\[6pt]
-&{{3g^4}\over{2m_W^2}}\left({{m_t^2}\over
{s_{\beta}^2}}+{{m_b^2}\over{c_{\beta}^2}}\right)
+{{3g^2m_t^2m_b^2}\over{s_{\beta}^2c_{\beta}^2m_W^4}}\Biggr]
\ln\left({{\msusy^2}\over{\mw^2}}\right)\,.
\label{lcuaunloop}
\eeqno
}

\noindent The terms proportional to the number of generations $N_g=3$
and the number of Higgs doublets $N_H=2$ that remain in the
low-energy effective theory at the scale $\mu=\mw$ have their origin in
the running of $g^2$ from $\msusy$ down to $\mw$.
In deriving this expression, I have taken $\mweak=\mw$.  This is
a somewhat arbitrary decision, since another reasonable choice would
yield a result that differs from eq.~(\ref{lcuaunloop}) by a non-leading
logarithmic term.  Comparisons with a more complete
calculation show that
one should choose $\mweak=\mw$ in computations involving the charged
Higgs (and gauge) sector, and $\mweak=\mz$ in computations involving the
neutral sector.

The above analysis also assumes that $m_t\sim {\cal O}(m_W)$.  Although
this is a good assumption, we can improve the above result somewhat
by decoupling the $(t,b)$ weak doublet from
the low-energy theory for scales below $m_t$.  The
terms in eq.~(\ref{lcuaunloop}) that are proportional to $m_t^2$ and/or
$m_b^2$ arise from self-energy diagrams containing a $tb$ loop.
Thus, such a term should not be present for $\mw\leq \mu\leq m_t$.
In addition, we recognize the term in eq.~(\ref{lcuaunloop}) proportional
to the number of generations $N_g$ as arising from the contributions
to the self-energy diagrams containing either quark or lepton loops
(and their supersymmetric partners).
To identify the contribution of the $tb$ loop to this term,
simply write
\vskip6pt
\beq
N_g=\quarter N_g(N_c+1)=\quarter N_c+\quarter[N_c(N_g-1)+N_g]\,,
\label{ngee}
\eeq
\vskip6pt \noindent%
where $N_c=3$ colors.  Thus, we identify $\quarter N_c$ as the piece
of the term proportional to $N_g$ that is due to the $tb$ loop.  The
rest of this term is then attributed to the lighter quarks and leptons.
Finally, the remaining terms in eq.~(\ref{lcuaunloop}) are due to the
contributions from the gauge and
Higgs boson sector.  The final result is \cite{madiaz}
\beqno
\lambda_4(\mw^2) &=-\half g^2
-{N_c g^4\over{32\pi^2}}\left[{1\over 3}
-{1\over{2m_W^2}}\biggl({{m_t^2}\over
{s_{\beta}^2}}+{{m_b^2}\over{c_{\beta}^2}}\biggr)
+{{m_t^2m_b^2}\over{s_{\beta}^2c_{\beta}^2m_W^4}}\right]
\ln\left({{\msusy^2}\over{m_t^2}}\right) \nonumber\\ [6pt]
&\quad-{1\over 96\pi^2}\left\{\left[N_c(N_g-1)+N_g
+\half N_H-10\right]g^4
+15g^2g'^2 \right\}\ln\left({\msusy^2\over\mw^2}\right)\, .  \nonumber\\
\label{lambdaiv}
\eeqno

\pagebreak\noindent
Inserting this result (and $\lambda_5=0$) into eq.~(\ref{mamthree}),
we obtain the one-loop leading-logarithmic
(1LL) formula for the charged Higgs mass
\vspace*{6pt}
\beqno%
(m_{H^{\pm}}^2)_{\rm 1LL}&=m_A^2+m_W^2 +{{N_c g^2}\over{32\pi^2m_W^2}}
\Bigg[{{2m_t^2m_b^2}\over{s_{\beta}^2c_{\beta}^2}}-m_W^2
\bigg({{m_t^2}\over{s_{\beta}^2}}+{{m_b^2}\over{c_{\beta}^2}}\bigg)
+{\textstyle{2\over 3}}m_W^4\Bigg]  \nonumber \\ [9pt]
&\times\ln\left({{\msusy^2}\over{m_t^2}}\right)
+{{m_W^2}\over{48\pi^2}} \left\{\left[N_c(N_g-1)+N_g
-9\right]g^2 +15g'^2\right\}
\ln\left({{\msusy^2}\over{m_W^2}}\right)\,.   \nonumber\\
\label{llform}
\eeqno
\vskip6pt \noindent%
Since this derivation makes use of the two-Higgs-doublet RGEs for
the $\lambda_i$, there is an implicit assumption that the full
two-doublet Higgs spectrum survives in the low-energy effective theory
at $\mu=\mw$.  Thus, I have set $N_H=2$ in obtaining eq.~(\ref{llform})
above.  It also means that $\mha$ cannot be much larger than
$\mw$.\footnote{If $\mha\sim {\cal O}(\msusy)$, then $H^\pm$, $\hh$ and
$\ha$ would all have masses of order $\msusy$, and the effective
low-energy theory below $\msusy$ would be that of the minimal
Standard Model.
For example, for $\mha=\msusy$, the leading logarithmic corrections
to the charged Higgs mass can be obtained from $\mhpm^2=\mha^2+\mw^2$
by treating $\mw^2$ as a running parameter evaluated at $\mha$.
Re-expressing $\mw(\mha)$ in terms of the physical $W$ mass yields
the correct one-loop leading log correction to $\mhpm^2$.  For
$\mz\leq\mha\leq\msusy$, one can interpolate between the effective
two-Higgs doublet model and the effective one-Higgs doublet model.}
The leading logarithms of eq.~(\ref{llform}) can be resummed to all
orders of perturbation theory by using the full RGE solution to
$\lambda_4(\mw^2)$
\vskip6pt
\beq
\mhpm^2=\mha^2-\half\lambda_4(\mw^2)(v_1^2+v_2^2)\,.\label{chiggsrge}
\eeq

\vspace*{1pc}
Although the one-loop leading-log formula for $\mhpm$
[eq.~(\ref{llform})]
gives a useful indication as to the size of the radiative corrections,
non-leading logarithmic contributions can also be important
in certain regions of parameter space.  A more complete set of
radiative corrections can be found in the literature
\cite{berz,turski,brig,madiaz,completeoneloop}.
However, it should be emphasized that the radiative corrections
to the charged Higgs mass are significant only for $\tanb<1$, a
region of MSSM parameter space not favored in supersymmetric
models.

The computation of the neutral CP-even Higgs masses follows a
similar procedure.  The results of Ref.~\cite{llog} are summarized
below.
From eq.~(\ref{massmhh}), we see that we only need results for $\lambda_1$,
$\lambda_2$ and $\widetilde\lambda_3\equiv\lambda_3+\lambda_4+\lambda_5$.
(Recall that $\lambda_5=\lambda_6=\lambda_7=0$ at all energy scales.)
By iterating the corresponding RGEs as before, we find

\pagebreak
\vbox{%
\beqno
\lambda_1(\mz^2)&=~~\fourth[g^2+g'^2](\mz^2)
+{g^4\over384\pi^2\cw^4}\Bigg[ P_t\ln\left({\msusyy\over
 m_t^2}\right)                                \nonumber\\
&\qquad+\bigg(12N_c{m_b^4\over\mz^4\cb^4}-6N_c{m_b^2\over\mzz\cb^2}
+P_b+P_f+P_g+P_{2H} \bigg)\ln\left({\msusyy\over
m_Z^2}\right)\Bigg]\,,    \nonumber \\[3pt]
\lambda_2(\mz^2)&=~~\fourth [g^2+g'^2](\mz^2)
+{g^4\over384\pi^2\cw^4}\Bigg[\bigg(P_b+P_f+P_g+P_{2H}
\bigg)\ln\left({\msusyy\over m_Z^2}\right)     \nonumber\\
&\qquad+\bigg(12N_c{m_t^4\over\mz^4\sb^4}-6N_c{m_t^2\over\mzz\sb^2}
+P_t\bigg)\ln\left({\msusyy\over m_t^2}\right)\Bigg]\,, \nonumber \\[3pt]
\widetilde\lambda_3(\mz^2)&=-\fourth[g^2+g'^2](\mz^2)
-{g^4\over384\pi^2\cw^4}\Bigg[\bigg(P_t-3N_c{m_t^2\over\mzz\sb^2}
\bigg)\ln\left({\msusyy\over m_t^2}\right) \nonumber\\
&\qquad+\bigg(-3N_c{m_b^2\over\mzz\cb^2}+P_b+P_f+P_g'+P_{2H}'
\bigg)\ln\left({\msusyy\over m_Z^2}\right)\Bigg]\,,\label{dlambda}
\eeqno
}

\noindent where
\beqno
P_t~&\equiv~~N_c(1-4e_t\sw^2+8e_t^2\sw^4)\,, \nonumber\\[3pt]
P_b~&\equiv~~N_c(1+4e_b\sw^2+8e_b^2\sw^4)\,, \nonumber\\[3pt]
P_f~&\equiv~~ N_c(N_g-1)[2-4\sw^2+8(e_t^2+e_b^2)\sw^4]
+N_g[2-4\sw^2+8\sw^4]\,,  \nonumber\\[3pt]
P_g~&\equiv-44+106\sww-62\sw^4\,, \nonumber \\[3pt]
P_g'~&\equiv~~10+34\sww-26\sw^4\,, \nonumber\\[3pt]
P_{2H}&\equiv -10+2\sww-2\sw^4\,, \nonumber\\ [3pt]
P_{2H}'&\equiv~~8-22\sww+10\sw^4\,.
\label{defpp}
\eeqno
\vskip6pt \noindent%
In the above formulae, the electric charges of the quarks are $e_t
= 2/3$, $e_b = -1/3$, and the subscripts $t, b, f, g$ and $2H$
indicate that these are the contributions from the
top and bottom quarks, the other fermions (leptons and the first two
generations of quarks),
the gauge bosons and the Higgs doublets, and the corresponding
supersymmetric partners, respectively.

As in the derivation of $\lambda_4(\mw^2)$ above, we have improved
our analysis by removing the effects of top-quark loops below
$\mu=\mt$.  This requires a careful treatment of the evolution of
$g$ and $g'$ at scales below $\mu=\mt$.  The correct procedure is
somewhat subtle, since the full electroweak gauge symmetry is
broken below top-quark threshold; for further details, see
Ref.~\cite{llog}.  However, the following pedestrian
technique works: consider the RGE for $g^2+g'^2$ valid for $\mu<\msusy$
\vskip6pt
\beq
{d\over dt}(g^2+g'^2)={1\over 96\pi^2}\left[\left(8g^4
+\fortythirds g'^4\right)N_g+(g^4+g'^4)N_H-44g^4\right]\,.\label{grge}
\eeq
\vskip6pt\noindent
This equation is used to run $g^2+g'^2$, which appears in eq.~(\ref{boundary}),
from $\msusy$ down to $\mz$.  As before, we identify the term
proportional to $N_g$ as corresponding to the fermion loops.
We can explicitly extract the $t$-quark contribution by noting that
\vskip6pt
\beqno
N_g\left(8g^4+\fortythirds g'^4\right)&=
{g^4N_g\over\cw^4}
\left[\sixtyfourthirds s_W^4-16 s_W^2+8\right]\nonumber \\[3pt]
&={g^4\over\cw^4}\biggl\{N_c[1+(N_g-1)](1-4e_t s_W^2+8e_t^2
       s_W^4)\nonumber \\[3pt]
&\qquad + N_c N_g(1+4e_b s_W^2+8e_b^2 s_W^4)
+N_g(2-4 s_W^2+8s_W^4)\biggr\}\,, \nonumber\\
\label{pedest}
\eeqno
\vskip6pt\noindent
where in the first line of the last expression, the term proportional
to 1 corresponds to the $t$-quark
contribution while the term proportional
to $N_g-1$ accounts for the $u$ and $c$-quarks; the second line contains
the contributions from the down-type quarks and leptons respectively.
Thus, iterating to one-loop,
\vskip6pt
\beqno%
[g^2+g'^2](\msusy^2)&=
[g^2+g'^2](\mz^2)+{g^4\over 96\pi^2 c_W^4}
\Biggl[P_t\ln\left({\msusy^2\over\mt^2}\right)\nonumber \\[3pt]
&\quad+\left[P_b+P_f+(\sw^4+\cw^4)N_H-44\cw^4\right]
\ln\left({\msusy^2\over\mz^2}
\right)\Biggr]\,.\label{gaugeiter}
\eeqno
\noindent\vskip6pt
Again, we take $N_H=2$, since the low-energy effective theory between
$\mz$ and $\msusy$ consists of the full two-Higgs doublet model.
Eq.~(\ref{gaugeiter}) was used in the derivation of eq.~(\ref{dlambda}).

We now return to the computation of the one-loop leading log
neutral CP-even Higgs squared-mass matrix.
The final step is to insert the expressions obtained in eq.~(\ref{dlambda})
into eq.~(\ref{massmhh}).  The resulting
matrix elements for the mass-squared matrix to one-loop leading
logarithmic accuracy are given by

\vbox{%
\beqno
(\calm_{11}^2)_{\rm 1LL}&=\mha^2\sb^2+m_Z^2\cb^2
+{g^2\mzz\cb^2\over96\pi^2\cw^2}\Bigg[
P_t~\ln\left({\msusyy\over m_t^2}\right)  \nonumber\\
&\quad+\bigg(12N_c{m_b^4\over\mz^4\cb^4}-6N_c{m_b^2\over\mzz\cb^2}
+P_b+P_f+P_g+P_{2H} \bigg)\ln\left({\msusyy\over
m_Z^2}\right)\Bigg] \nonumber \\[3pt]
(\calm_{22}^2)_{\rm 1LL}&=\mha^2\cb^2+m_Z^2\sb^2
+{g^2\mzz\sb^2\over96\pi^2\cw^2}\Bigg[\bigg(P_b+P_f+P_g+P_{2H}
\bigg)\ln\left({\msusyy\over m_Z^2}\right)\nonumber \\[3pt]
&\quad+\bigg(12N_c{m_t^4\over\mz^4\sb^4}-6N_c{m_t^2\over\mzz\sb^2}
+P_t\bigg)\ln\left({\msusyy\over m_t^2}\right)\Bigg]
\nonumber\\[3pt]
(\calm_{12}^2)_{\rm 1LL}&=-\sb\cb\Biggl\{\mha^2+m_Z^2
+{g^2\mzz\over96\pi^2\cw^2}\Bigg[\bigg(P_t-3N_c{m_t^2\over\mzz\sb^2}
\bigg)\ln\left({\msusyy\over m_t^2}\right)\nonumber \\[3pt]
&\quad+\bigg(-3N_c{m_b^2\over\mzz\cb^2}+P_b+P_f+P_g'+P_{2H}'
\bigg)\ln\left({\msusyy\over
m_Z^2}\right)\Bigg]\Biggr\}\,,  \label{mtophree}
\eeqno
}

\noindent Diagonalizing this matrix [eq.~(\ref{mtophree})] yields
the radiatively corrected CP-even Higgs masses and mixing angle $\alpha$.

The analysis presented above assumes that $\mha$ is not much larger
than ${\cal O}(\mz)$ so that the Higgs sector of the
low-energy effective theory contains the full two-Higgs-doublet spectrum.
On the
other hand, if $\mha\gg\mz$, then only $\hl$ remains in the low-energy
theory.  In this case, we must integrate
out the heavy Higgs doublet, in which case
one of the mass eigenvalues of ${\cal M}_0^2$ [eq.~(\ref{kv})] is much
larger than the weak scale. In order to obtain the effective
Lagrangian at $\mweak$, we first have to run the various coupling
constants to the threshold $\mha$. Then we diagonalize the
Higgs mass matrix and express the Lagrangian in terms of the mass
eigenstates. Notice that in this case the mass eigenstate $\hl$ is
directly related to the field with the non-zero vacuum expectation value
[\ie, $\beta(\mha)=\alpha(\mha)+\pi/2+{\cal O}(\mz^2/\mha^2)$].

Below $\mha$ only the Standard Model Higgs doublet
$\phi\equiv\cb\Phi_1+\sb\Phi_2$ remains. The scalar potential is
\beq
{\cal V}=m_{\phi}^2(\phi^{\dagger}\phi)
+\half\lambda(\phi^{\dagger}\phi)^2
\,,\label{potsm}
\eeq
and the light CP-even Higgs mass is obtained using
$\mhl^2=\lambda v^2$.
The RGE in the Standard Model for $\lambda$ is \cite{chengli,cmpp}
\beq
16\pi^2\beta_{\lambda} = 6\lambda^2
+\threeighth \left[2g^4+(g^2+g'^2)^2\right]-2\sum_i
N_{c_i}h_{f_i}^4 -\lambda\biggl(\ninehalf
g^2+\threehalf g'^2-2\sum_i N_{c_i}
h_{f_i}^2\biggr),
\label{defbetl}
\eeq
\vskip12pt\noindent
where the summation is over all fermions with
$h_{f_i}=gm_{f_i}/(\sqrt{2}\mw)$. The RGEs for the gauge couplings
are obtained from $\beta_{g^2}$ and $\beta_{g'^2}$
given in Appendix B by putting
$N_H=1$.  In addition, we require
the boundary condition for $\lambda$ at $\mha$
\beqno
\lambda(\mha)&=\left[\cb^4\lambda_1+\sb^4\lambda_2+
2\sb^2\cb^2(\lambda_3+\lambda_4+\lambda_5)
+4\cb^3\sb\lambda_6+4\cb\sb^3\lambda_7\right](\mha)\nonumber \\[3pt]
&=\left[\fourth(g^2+g'^2)\ctwob^2\right](\mha)
+{g^4\over384\pi^2\cw^4} \ln\left({\msusyy\over\mha^2}\right)\nonumber \\[3pt]
&~\times\bigg[12N_c\bigg({m_t^4\over\mz^4}+{m_b^4\over\mz^4}\bigg)
+6N_c\ctwob\bigg({m_t^2\over\mzz}-{m_b^2\over\mzz}\bigg) \nonumber\\
&~+\ctwob^2\big(P_t+P_b+P_f\big)+(\sb^4+\cb^4)(P_g+P_{2H})
-2\sb^2\cb^2(P_g'+P_{2H}')\bigg]\,,
 \label{lapprox}
\eeqno
\vskip12pt\noindent
where $(g^2+g'^2)c_{2\beta}^2$ is to be evaluated at the scale
$\mha$ as indicated.
The RGE for $g^2+g'^2$ was given in eq.~(\ref{grge});
note that at scales below $m_A$ we must set $N_H=1$.

Finally,
we must deal with implicit scale dependence of $c_{2\beta}^2$.
Since the fields $\Phi_i$ $(i=1,2)$ change with the scale, it
follows that $\tanb$ scales like the ratio of the two Higgs doublet
fields, \ie,
\beq
{1\over\tan^2\beta}{d\tan^2\beta\over
dt}={\Phi_1^2\over\Phi_2^2}{d\over dt}
\left({\Phi_2^2\over\Phi_1^2}\right)
=\gamma_2-\gamma_1\,.\label{deftanb}
\eeq
Thus we arrive at the RGE for $\cos 2\beta$ in terms of the
anomalous dimensions $\gamma_i$ given in eq.~(\ref{wavez}).
Solving this equation iteratively to first order yields
\beq
\ctwob^2(\mha)=\ctwob^2(\mz)
+4\ctwob\cb^2\sb^2(\gamma_1-\gamma_2)
\ln\left({\mha^2\over\mzz}\right)\,.\label{rgecb}
\eeq

The one loop leading log expression for $\mhl^2 = \lambda(\mz)
v^2$ can now be obtained by solving the RGEs above for $\lambda(\mz)$
iteratively to first order using the boundary condition given
in eq.~(\ref{lapprox}).  The result is
\beqno
(\mhl^2)_{\rm 1LL}&= \mzz\ctwob^2(\mz)
+{g^2m_Z^2\over96\pi^2\cw^2}
\Bigg\{\bigg[12N_c{m_b^4\over\mz^4}-6N_c\ctwob{m_b^2\over\mzz}
+\ctwob^2(P_b+P_f)  \nonumber\\
&~~+\left(P_{g}+P_{2H})(\sb^4+\cb^4\right)
 -2\sb^2\cb^2\left(P_{g}'+P_{2H}'\right)
 \bigg]\ln\left({\msusyy\over\mzz} \right) \nonumber\\
&~~+\bigg[12N_c{m_t^4\over\mz^4}+
 6N_c\ctwob{m_t^2\over\mzz}+\ctwob^2P_t\bigg]
 \ln\left({\msusyy\over m_t ^2}\right)  \nonumber\\
&~~-\bigg[\left(\cb^4+\sb^4\right)P_{2H}-2\cb^2\sb^2P_{2H}'-P_{1H}\bigg]
 \ln\left({\mha^2\over\mzz}\right)\Bigg\} \,, \label{mhltot}
\eeqno
where the term proportional to
\beq
P_{1H} \equiv -9\ctwob^4+(1-2\sww+2\sw^4)\ctwob^2\,,
\label{defpps}
\eeq
corresponds to the Higgs
boson contribution in the one-Higgs-doublet model.
The term in eq.~(\ref{mhltot}) proportional to $\ln(\mha^2)$
accounts for the fact that there are two Higgs doublets present at
a scale above $\mha$ but only one Higgs doublet below $\mha$.

We can improve the above one-loop leading log
formulae by reinterpreting the meaning of
$\msusy$.  For example, all terms proportional to $\ln(\msusy^2/\mt^2)$
arise from diagrams with loops involving the top quark and top-squarks.
Explicit diagrammatic computations then show that
we can reinterpret $\msusy^2=M_{\tilde t_1}M_{\tilde t_2}$.
Note that with this reinterpretation of $\msusyy$, the top quark and top
squark loop contributions to the Higgs masses cancel exactly
when $M_{\tilde t_1}=M_{\tilde t_2}=m_t$, as required
in the supersymmetric limit.
Likewise, in terms proportional to $P_b$ or powers of $m_b$ multiplied by
$\ln(\msusyy/\mzz)$, we may reinterpret $\msusy=M_{\tilde b_1}M_{\tilde b_2}$.
Terms proportional to $P_f\ln(\msusyy/\mzz)$
come from loops of lighter quarks and
leptons (and their supersymmetric partners) in an obvious way, and
the corresponding $\msusyy$ can be reinterpreted accordingly.
The remaining leading logarithmic terms arise from gauge and
Higgs boson loops and their supersymmetric partners.  The best we can
do in the above formulae is to interpret $\msusy$ as an average
neutralino and chargino mass.  To incorporate thresholds more precisely
requires a more complicated version of eq.~(\ref{mtophree}),
which can be easily derived from formulae given in Ref.~\cite{llog}.
The explicit form of these threshold corrections can be found in
Ref.~\cite{hhh}.  However,
the impact of these corrections are no more important than the
non-leading logarithmic terms which have been discarded.

The largest of the non-leading logarithmic terms is
of ${\cal O}(g^2\mt^2)$, which can be identified from a full one-loop
computation as being the subdominant term relative to the leading
${\cal O}(g^2\mt^4\ln\msusy^2)$ term in ${\cal M}_{22}^2$.  Thus,
we can make a minor improvement on our computation of the
one-loop leading-log
CP-even Higgs squared mass matrix by taking
\beq
{\cal M}^2 = {\cal M}^2_{\rm 1LL} +
{N_c g^2\mt^2\over48\pi^2\sb^2c_W^2}\left(
\begin{array}{cc} 0&0\\0&1 \end{array}\right)\,.
\label{nllogtrm}
\eeq
where $\calm^2_{\rm 1LL}$ is the matrix whose elements are given in
eq.~(\ref{mtophree}).
One can check that this yields at most a 1~GeV shift in
the computed Higgs masses.

\subsection{Leading Squark Mixing Corrections to the MSSM Higgs Masses}

In the case of multiple and widely separated
supersymmetric particle thresholds and/or
large squark mixing (which is most likely in the top squark sector),
new non-leading logarithmic contributions to the scalar mass-squared
matrix can become important.  As shown in Ref.~\cite{llog},
such effects can be taken into account by modifying the
boundary conditions of the $\lambda_i$ at the supersymmetry breaking
scale [eq.~(\ref{boundary})], and by
modifying the RGEs to account for multiple thresholds.  In
particular, we find that $\lambda_5$, $\lambda_6$ and $\lambda_7$
are no longer zero. If the new RGEs are solved iteratively to one
loop, then the effects of the new boundary conditions are simply
additive.

In this section, we focus on the
effects arising from the mass splittings and $\widetilde
q_L$--$\widetilde q_R$ mixing
in the third generation
squark sector.  The latter generates additional squared-mass shifts
proportional to $m_t^4$ and thus can
have a significant impact on the radiatively
corrected Higgs masses \cite{erz}.
First, we define our notation (we follow the
conventions of Ref.~\cite{hehtasi}).  In third family notation, the
squark
mass eigenstates are obtained by diagonalizing the following two
$2\times 2$ matrices.  The top-squark squared-masses are eigenvalues of
\begin{equation}
\left(\begin{array}{cc}
  M_{Q}^2+m_t^2+t_L m_Z^2 & m_t X_t \\
  m_t X_t & M_{U}^2+m_t^2+t_R m_Z^2
\end{array}\right)  \,,
\label{stopmatrix}
\end{equation}
where $X_t \equiv A_t-\mu\cot\beta$,
$t_L\equiv ({1\over 2}-e_t\sin^2\theta_W)\cos2\beta$ and
$t_R\equiv e_t\sin^2\theta_W\cos2\beta$.
The bottom-squark squared-masses are eigenvalues of
\begin{equation}
\left(\begin{array}{cc}
  M_{Q}^2+m_b^2+b_L m_Z^2 & m_b X_b \\
 m_b X_b & M_{D}^2+m_b^2+b_R m_Z^2
\end{array}\right) \,,
\label{sbotmatrix}
\end{equation}
where $X_b \equiv A_b -\mu\tan\beta$,
$b_L\equiv (-{1\over 2}-e_b\sin^2\theta_W)\cos2\beta$ and
$b_R\equiv e_b\sin^2\theta_W\cos2\beta$.
$M_{Q}$, $M_{U}$, $M_{D}$, $A_t$, and
$A_b$ are soft-supersymmetry-breaking parameters,
and $\mu$ is the supersymmetric Higgs mass parameter.
We treat the squark mixing perturbatively, assuming that the off-diagonal
mixing terms are small compared to the diagonal terms.

At one-loop, the effect of the
squark mixing is to introduce the shifts $\Delta \calmm_{\rm mix}$
and $\left(\Delta\mhpm^2\right)_{\rm mix}$.
In order to keep the formulae simple,
we take $M_Q=M_U=M_D=\msusy$, where $\msusy$ is assumed to be
large compared to $\mz$.  Thus, the radiatively corrected Higgs
mass is determined by $\mha$, $\tanb$, $\msusy$, $A_t$, $A_b$, and
$\mu$.  The
more complex case of non-universal squark squared-masses
(in which $M_Q$, $M_U$, and $M_D$ are unequal but still large compared
to $\mz$) is treated in Ref.~\cite{hhh}.

It is convenient to define
\begin{eqnarray}
X_t&\equiv&A_t-\mu\cot\beta\,,\qquad\qquad
Y_t\equiv A_t+\mu\tan\beta\,,\nonumber \\
X_b&\equiv&A_b-\mu\tan\beta\,,\qquad\qquad
Y_b\equiv A_b+\mu\cot\beta\,.
\label{xdefs}
\end{eqnarray}
We assume that the mixing terms $\mt X_t$ and $\mb X_b$ are not too
large.\footnote{Formally, the expressions given in
eqs.~(\ref{fullcorr})--(\ref{deltamhpm})
are the results of an expansion in the variable $(M_1^2-M_2^2)/
(M_1^2+M_2^2)$, where $M_1^2$, $M_2^2$ are the squared-mass eigenvalues
of the squark mass matrix.   Thus, we demand that $\mt X_t/\msusyy\ll
1$.  For example, for $\msusy=1$~TeV, values of $X_t/\msusy\lsim 3$
should yield an acceptable approximation based on the formulae
presented here.}
Then, the elements of the CP-even Higgs squared-mass matrix
are given by:
\begin{equation} \label{fullcorr}
\calmm=\calmm_{\rm 1LL}+\Delta\calmm_{\rm mix}\,,
\end{equation}
where $\calmm_{\rm 1LL}$ has been given in
eq.~(\ref{mtophree}), and
\begin{eqnarray} \label{deltacalms}
&&(\Delta\calmm_{11})_{\rm mix} =  {g^2N_c\over 32\pi^2\mw^2\msusy^2}\Biggl[
{4m_b^4A_b X_b\over\cbb}\left(1-{A_b X_b\over 12\msusy^2}\right)
-{m_t^4\mu^2 X_t^2\over 3\msusy^2\sbb}\nonumber \\
  &&\qquad -\mz^2 m_b^2A_b(X_b+\third A_b)-\mz^2
m_t^2\mu\cot\beta(X_t+\third\mu\cot\beta)\Biggr]\,,
\nonumber \\
&&(\Delta\calmm_{22})_{\rm mix} =  {g^2N_c\over 32\pi^2\mw^2\msusy^2}\Biggl[
  {4m_t^4A_t X_t\over\sbb}\left(1-{A_t X_t\over 12\msusy^2}\right)
  -{m_b^4\mu^2 X_b^2\over 3\msusy^2\cbb}\nonumber \\
  &&\qquad -\mz^2 m_t^2A_t(X_t+\third A_t)-\mz^2
  m_b^2\mu\tan\beta(X_b+\third\mu\tan\beta)\Biggr]\,,
\nonumber \\
&&(\Delta\calmm_{12})_{\rm mix} =  {-g^2N_c\over 64\pi^2\mw^2\msusy^2}
  \Biggl[ {4m_t^4\mu X_t\over\sbb} \left(\!1-{A_t X_t\over 6\msusy^2}\right)
  +{4m_b^4\mu X_b\over\cbb}\left(\!1-{A_b X_b\over 6\msusy^2}\right)\nonumber \\
  &&\qquad-\mz^2 m_t^2\cot\beta\left[X_t Y_t+\third(\mu^2+A_t^2)\right]
  -\mz^2 m_b^2\tan\beta\left[X_b Y_b+\third(\mu^2+A_b^2)\right]\Biggr]\,.
\end{eqnarray}

If $\mz\ll\mha\leq\msusy$, a separate analysis is required.  One finds
that
eq.~(\ref{mhltot}) is shifted by
\begin{eqnarray} \label{deltamhs}
&&(\Delta\mhl^2)_{\rm mix}\!=\!{g^2 N_c\over 16\pi^2\mw^2\msusyy}
\Biggl\{2m_t^4 X_t^2\left(\!1-{X_t^2\over
12\msusyy}\right)+2m_b^4 X_b^2\left(\!1-{X_b^2\over
12\msusyy}\right) \nonumber \\
&&+\half\mz^2\cos2\beta\left[m_t^2
\left(X_t^2+\third(A_t^2\!-\!\mu^2\cot^2\beta)\right)
\!-m_b^2\left(X_b^2+\third(A_b^2\!-\!\mu^2\tan^2\beta)\right)\right]
\!\Biggr\}.
\end{eqnarray}

Squark mixing effects
also lead to modifications of the charged Higgs squared-mass.
One finds that the charged Higgs squared-mass
obtained in eq.~(\ref{llform}) is shifted by

\vbox{%
\begin{eqnarray} \label{deltamhpm}
&&(\mhpm^2)_{\rm mix} =
 {N_cg^2\over192\pi^2\mw^2\msusy^2}\Biggl[
 {2\mt^2\mw^2(\mu^2-2A_t^2)\over\sbb}
  +{2\mb^2\mw^2(\mu^2-2A_b^2)\over\cbb}\hspace*{1cm} \nonumber \\
&&\quad -3\mu^2\left({m_t^2\over\sbb}
  +{m_b^2\over\cbb}\right)^2 +{\mt^2\mb^2\over\sbb\cbb}\left(3(A_t+A_b)^2
-{(A_t A_b -\mu^2)^2\over\msusyy}\right)\Biggr].
\end{eqnarray}
}

\section{RG-Improvement and Numerical Results for the MSSM Higgs Masses}
\label{sec:five}

The RG-improved Higgs masses (in the absence of squark mixing)
are computed by solving the set of coupled
REGs for the $\lambda_i(\mweak^2)$, subject to the boundary conditions
specified in eq.~(\ref{boundary}).  Squark mixing effects are
incorporated into the procedure by modifying the boundary conditions as
described in Ref.~\cite{llog}.  Hempfling, Hoang
and I \cite{hhh} found a
simple analytic algorithm which reproduces quite accurately the results
of the numerical integration of the RGEs.

The procedure starts with the formulae of section 4.
The Higgs masses take the form given symbolically in
eq.~(\ref{oneloopmasses}). Then,
\begin{equation}
\calmm_{\rm 1RG}\simeq\overline{\calmm}_{\rm 1LL}+
\Delta\overline{\calmm}_{\rm mix}\equiv
\calmm_{\rm 1LL}\left[m_t(\mu_t),m_b(\mu_b)\right]+
\Delta\calmm_{\rm mix}\left[m_t(\mu_{\tilde t}),m_b(\mu_{\tilde b})\right]\,,
\label{simplemixform}
\end{equation}
where
\beq \label{scales}
 \mu_t\equiv\sqrt{m_t\msusy}\,,\qquad   \mu_b\equiv\sqrt{m_Z\msusy}\,,
\qquad \mu_{\tilde q}\equiv\msusy~~~(q=t,b)\,.
\eeq
 That is, the numerically integrated
RG-improved CP-even Higgs
squared-mass matrix, $\calmm_{\rm 1RG}$,
is well approximated by replacing all
occurrences of $m_t$ and $m_b$ in $\calmm_{\rm 1LL}(m_t,m_b)$
and $\Delta\calmm_{\rm mix}(m_t,m_b)$ by
the corresponding running masses evaluated at the scales as indicated
above.\footnote{In this section, an overline
above a quantity will indicate that the replacement of $m_t$ and $m_b$
by the appropriate running mass has been made.}
To implement the above algorithm, we need formulae for $m_b(\mu)$ and
$m_t(\mu)$.  First, consider $\mha={\cal O}(\mz)$.  In this case,
at mass scales below $\msusy$, the effective theory of the Higgs
sector is that of a non-supersymmetric two-Higgs-doublet model.
In this model, the
quark mass is the product of the Higgs-quark Yukawa coupling
($h_q$) and the appropriate Higgs vacuum expectation value:
\begin{eqnarray}
m_b(\mu) & = & \frac{1}{\sqrt{2}}\,h_b(\mu)\,v_1(\mu)\,,\nonumber \\
m_t(\mu) & = & \frac{1}{\sqrt{2}}\,h_t(\mu)\,v_2(\mu)\,.
\label{topmass}
\end{eqnarray}
At scales $\mu\leq\msusy$, we employ the one-loop non-supersymmetric
RGEs of the two-Higgs doublet model for $h_b$, $h_t$,
and the vacuum expectation values $v_1$ and $v_2$ (see Appendix B).
This yields
\begin{eqnarray}
&&\frac{\rm d}{\rm d\ln\mu^2}\,m_b^2 =
\frac{1}{64\,\pi^2}\,\left[\,6 h_b^2+2 h_t^2-32 g_s^2
+\frac{4}{3} g^{\prime 2}\,\right]\,m_b^2\,, \nonumber \\
&&\frac{\rm d}{\rm d\ln\mu^2}\,m_t^2 =
\frac{1}{64\,\pi^2}\,\left[\,6 h_t^2+2 h_b^2-32 g_s^2
-\frac{8}{3} g^{\prime 2}\,\right]\,m_t^2\,.
\label{mtrge}
\end{eqnarray}
For $\mha={\cal O}(\msusy)$, the effective theory of the Higgs
sector at mass scales below $\msusy$
is that of the one-Higgs doublet Standard Model.  In this case,
we define $m_q(\mu)=h_q^{\rm SM}(\mu) v(\mu)/\sqrt{2}$, where
$v(\mz)\simeq 246$~GeV is the
one-Higgs-doublet Standard Model vacuum expectation value.  In this
case eq.~(\ref{mtrge}) is modified by replacing $6h_t^2+2h_b^2$
with $6(h_t^{\rm SM})^2-6(h_b^{\rm SM})^2$ in the RGE for $m_t^2$
(and interchange $b$ and $t$ to obtain the RGE for $m_b^2$).

To solve these equations, we also need the evolution equations of
$g_s$, and $g^\prime$.  But, an approximate solution is
sufficient for our purposes.  Since $g^\prime$ is small, we drop it.
We do not neglect the $h_b$ dependence
which may be significant if $\tan\beta$ is large.
Then, we can iteratively solve eq.~(\ref{mtrge}) to one loop by
ignoring the $\mu$ dependence of the right hand side.  We find
\begin{equation}
m_t(\mu) = m_t(m_t)\times
\begin{cases}1-{1\over\pi}\left[\alpha_s-{1\over 16}
(\alpha_b+3\alpha_t)\right]\,
\ln\left(\mu^2/m_t^2\right)\,, & $\mha\simeq{\cal
O}(\mz)\,,$\\
1-{1\over\pi}\left[\alpha_s-{3\over 16}
(\alpha_t^{\rm SM}-\alpha_b^{\rm SM})\right]\,
\ln\left(\mu^2/m_t^2\right)\,, & $\mha\simeq{\cal
O}(\msusy)\,,$  \end{cases}
\label{mtrun}
\end{equation}
where $\alpha_t\equiv h_t^2/4\pi$, {\it etc.}, and
all coupling on the right hand side are evaluated at $m_t$.
Similarly,
\begin{equation}
m_b(\mu) = m_b(\mz)\times
\begin{cases}1-{1\over\pi}\left[\alpha_s-{1\over 16}
(\alpha_t+3\alpha_b)\right]\,
\ln\!\left(\mu^2/\mz^2\right),&$\mha\simeq{\cal
O}(\mz),$ \\
1-{1\over\pi}\left[\alpha_s-{3\over 16}
(\alpha_b^{\rm SM}\!-\alpha_t^{\rm SM})\right]\,
\ln\!\left(\mu^2/\mz^2\right),&$\mha\simeq{\cal
O}(\msusy),$  \end{cases}
\label{mbrun}
\end{equation}
For intermediate values of $\mha$, one may
interpolate the above formulae between the two regions.
Using eqs.~(\ref{mtrun}) and (\ref{mbrun}) in
eq.~(\ref{simplemixform}), and
diagonalizing the resulting squared-mass matrix
yields our approximation to the RG-improved one-loop
neutral CP-even Higgs squared-masses.

We may also apply our algorithm to the radiatively corrected charged
Higgs mass.  However, in
contrast to the one-loop radiatively corrected neutral Higgs mass,
there are no one-loop leading logarithmic corrections to $\mhpm^2$ that
are proportional to $m_t^4$.  Thus, we expect that our charged Higgs
mass approximation will not be quite as reliable as our neutral Higgs mass
approximation.

Let us now compare various computations of the one-loop corrected
light CP-even Higgs mass. In the first set of examples, all squark
mixing effects are ignored.
First, we evaluate two expressions for the
RG-unimproved
one-loop Higgs mass---the one-loop leading log Higgs mass calculated
from $\calmm_{\rm 1LL}$ and from a simplified version of $\calmm_{\rm 1LL}$
in which only the dominant
terms proportional to $m_t^4$ are kept.  In the latter
case, we denote the neutral CP-even Higgs squared-mass matrix by
$\calmm_{\rm 1LT}\equiv \calmm_0+\Delta\calmm_{\rm 1LT}$, where
\begin{equation}
\Delta\calmm_{\rm 1LT}\equiv {3g^2 m_t^4\over 8\pi^2 m_W^2 \sbb}
 \ln\left(\msusyy/\mt^2\right)
\left(\begin{array}{cc}
  0 & 0 \\
  0 & 1
\end{array}\right)\,.
\label{topapprox}
\end{equation}

In many analyses of $\calmm_{\rm 1LT}$ and $\calmm_{\rm 1LL}$
that have appeared previously
in the literature, the Higgs mass radiative corrections were evaluated with
the pole mass, $m_t$.   Some have argued that one should take $m_t$ to be
the running mass evaluated at $m_t$, although to one-loop accuracy, the two
choices cannot be distinguished.  Nevertheless, because the leading
radiative effect is proportional to $m_t^4$, the choice of $m_t$ in the
one-loop formulae is numerically significant, and can lead to differences
as large as 10 GeV in the computed Higgs mass.
In Ref.~\cite{hhh},
the choice of using $m_t(m_t)$ as opposed to $m_t^{\rm pole}$
(prior to RG-improvement) is justified
by invoking information from a two-loop analysis.
Thus, our numerical results for the light CP-even Higgs mass before
RG-improvement are significantly lower (when $\msusy$ is large)
as compared to the original computations given in the literature, for fixed
$m_t^{\rm pole}$.  
We have taken $m_t(m_t)=166.5$~GeV in all the numerical results 
exhibited below.
We then apply our algorithm for RG-improvement by replacing $m_t$ and
$m_b$ by the appropriate running masses as
specified in eqs.~(\ref{simplemixform})--(\ref{scales}).

We now show examples for $\mha=1$~TeV and two choices of $\tanb$
in Fig.~\ref{hhhfig1} [$\tanb=20$]
and Fig.~\ref{hhhfig2} [$\tanb=1.5$], and for $\mha=100$~GeV and
$\tanb=20$ in Fig.~\ref{hhhfig3}.\footnote{For $\mha=100$~GeV and $\tanb=1.5$,
the resulting light Higgs mass lies below experimental Higgs mass bounds
obtained by the LEP collaborations \cite{LEPHIGGS}.}
Each plot displays five predictions for $\mhl$ based on
the following methods for computing the Higgs squared-mass matrix:
(i)~$\calmm_{\rm 1LT}$; (ii)~$\calmm_{\rm 1LL}$;
(iii)~$\overline{\calmm}_{\rm 1LT}$; (iv)~$\overline{\calmm}_{\rm 1LL}$;
and (v)~$\calmm_{\rm 1RG}$ [the overline notation is defined in the
footnote below eq.~(\ref{scales})].
The following general features are
noteworthy.  First, we observe that over the region of $\msusy$ shown,
$\calmm_{\rm 1RG}\simeq\overline{\calmm}_{\rm 1LL}$.  In fact, $\mhl$
computed from $\overline{\calmm}_{\rm 1LL}$ is within 1 GeV of the
numerical RG-improved $\mhl$ in all sensible
regions of the parameter space
($1\leq\tan\beta\leq m_t/m_b$ and $m_t$, $\mha\leq\msusy
\leq 2$~TeV).  For values of $\msusy>2$~TeV, the Higgs masses obtained
from $\overline{\calmm}_{\rm 1LL}$ begin
to deviate from the numerically integrated RG-improved result.
Second, the difference between $\mhl$ computed from $\calmm_{\rm 1LL}$
and from $\calmm_{1RG}$ is non-negligible for large values of $\msusy$;
neglecting RG-improvement can lead to an overestimate of $\mhl$ which
in some areas of parameter space can be as much as 10 GeV. Finally,
note that while the simplest approximation of $\mhl$ based on
$\calmm_{\rm 1LT}$ reflects the dominant radiative corrections, it yields
the largest overestimate of the light Higgs boson mass.

\begin{figure}[htb]
\centerline{\psfig{file=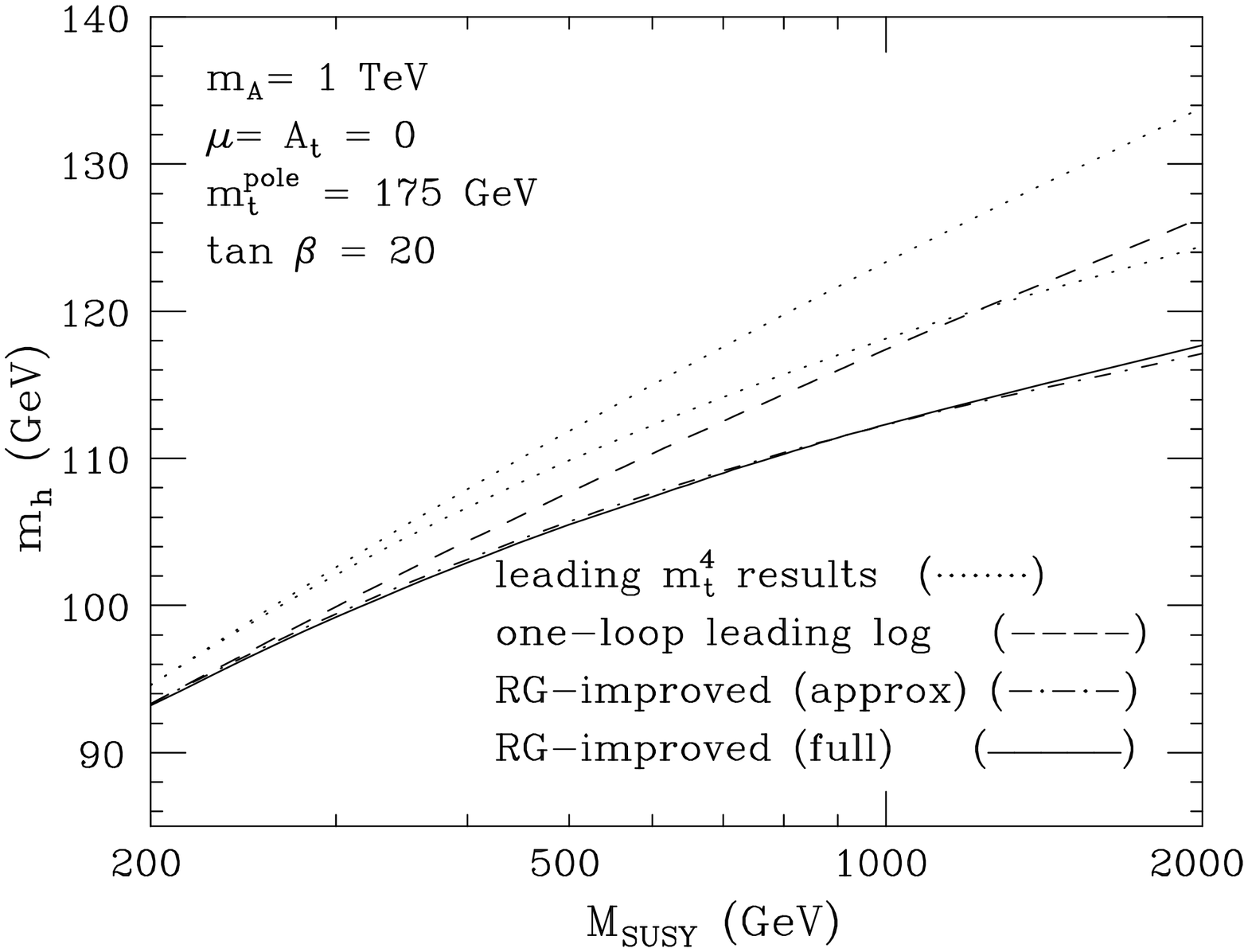,width=10cm,angle=90}}
\vskip1pc
\fcaption{The radiatively corrected light CP-even Higgs mass is plotted
as a function of $\msusy$ for $\tan\beta=20$ and $\mha= 1$~TeV.
The one-loop leading logarithmic computation [dashed line]
is compared with the RG-improved result which was obtained
by numerical analysis [solid line] and by using the simple analytic
result given in eq.~(\protect\ref{simplemixform}) [dot-dashed line].
For comparison, the
results obtained using the leading $m_t^4$ approximation of
eq.~(\protect\ref{topapprox}) [higher dotted line], and its RG-improvement
[lower dotted line] are also exhibited.  $\msusy$ characterizes the
scale of supersymmetry breaking and can be regarded (approximately)
as a common supersymmetric scalar mass; squark mixing effects are
set to zero.  The running top quark mass used in
our numerical computations is $m_t(m_t)= 166.5$~GeV.
All figures are taken from Ref.~\protect\cite{hhh}.}
\label{hhhfig1}
\end{figure}

\begin{figure}[hp]
\centerline{\psfig{file=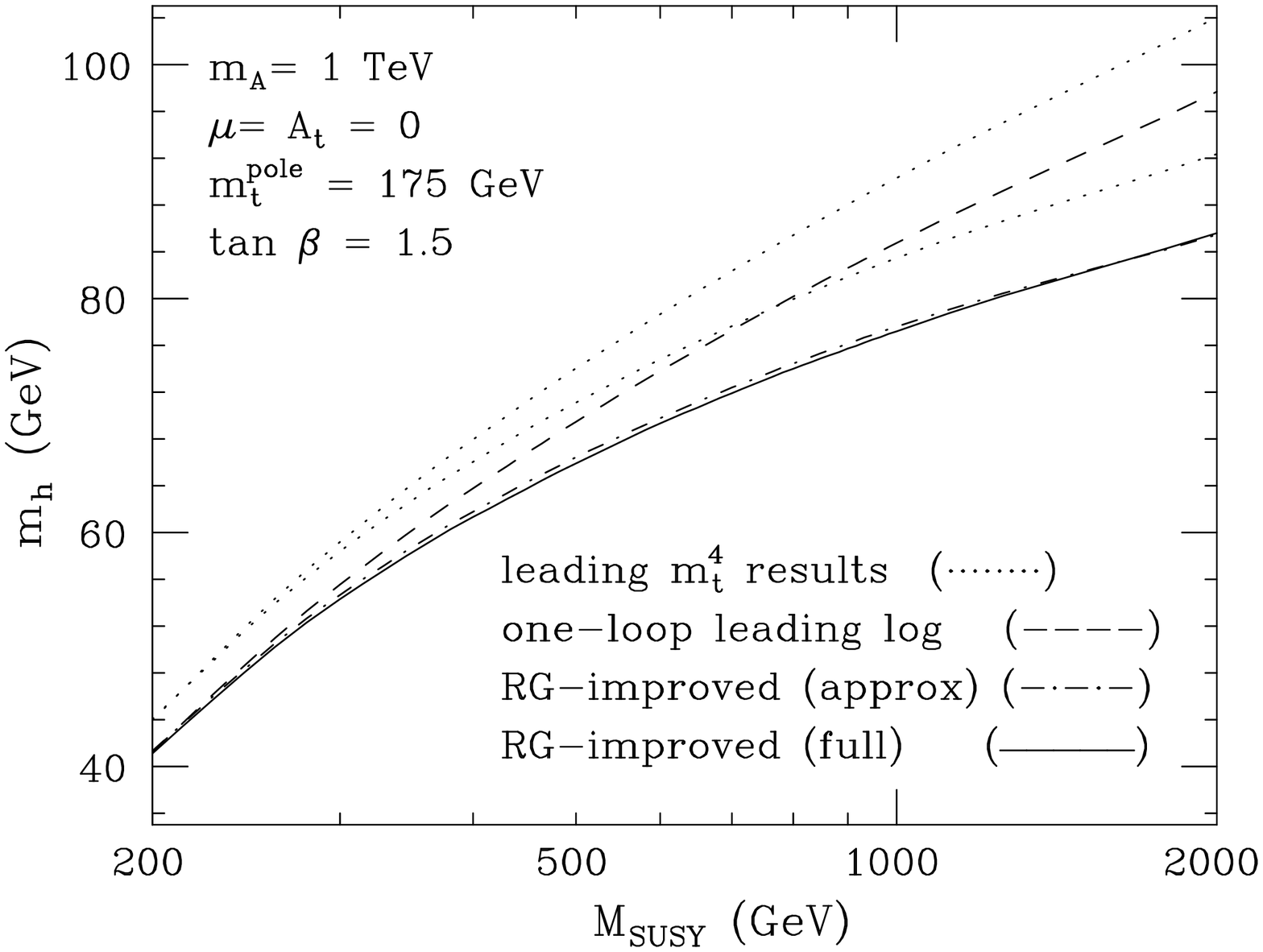,width=10cm,angle=90}}
\vskip1pc
\fcaption{The radiatively corrected light CP-even Higgs mass is plotted
as a function of $\msusy$ for $\tan\beta=1.5$ and $\mha= 1$~TeV.
See the caption to Fig.~\protect\ref{hhhfig1}.}
\label{hhhfig2}
\vspace*{1pc}
%
\centerline{\psfig{file=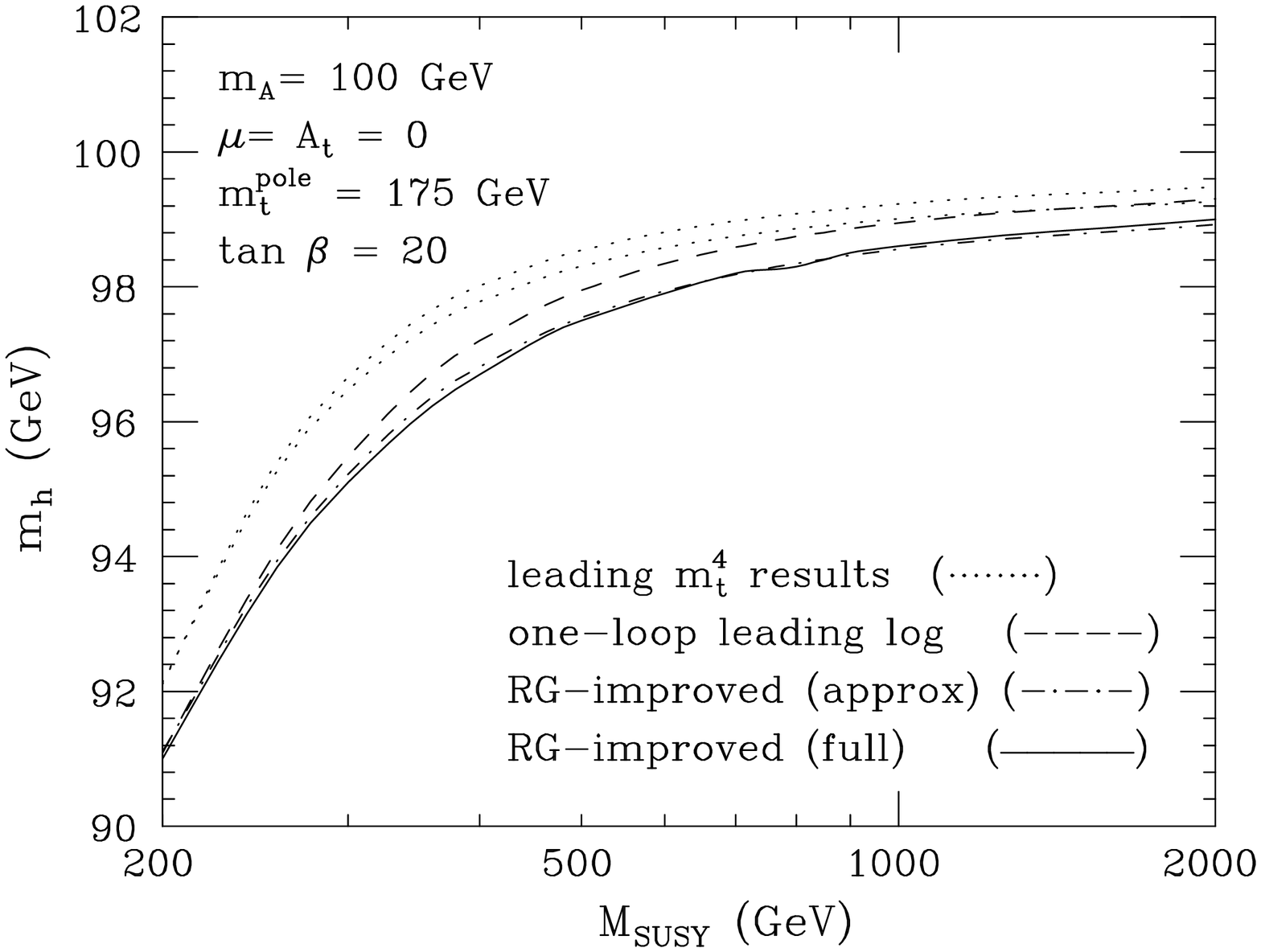,width=10cm,angle=90}}
\vskip1pc
\fcaption{The radiatively corrected light CP-even Higgs mass is plotted
as a function of $\msusy$ for $\tan\beta=20$ and $\mha= 100$~GeV.
See the caption to Fig.~\protect\ref{hhhfig1}.}
\label{hhhfig3}
\end{figure}

We next consider some examples in which squark-mixing effects are
included.  As above, we compare the value of
$\mhl$ computed by different procedures.  Prior to RG-improvement, we
first compute $\mhl$ by diagonalizing
$\calmm_{\rm 1LL}+\Delta\calmm_{\rm mix}$.
Next, we perform RG-improvement as in Ref.~\cite{llog}\
by numerically integrating the RGEs for the Higgs self-couplings
and inserting the results into 
eq.~(\ref{massmhh}); the resulting CP-even scalar
squared-mass matrix is denoted by
$\calmm_{\rm 1RG}$.
Finally, we extract $\mhl$ and compare it to the 
corresponding result obtained
by diagonalizing
$\overline{\calmm}_{\rm 1LL}+ \Delta\overline{\calmm}_{\rm mix}$ given by
eq.~(\ref{simplemixform}).  These comparisons
are exhibited in a series of figures.  First, we plot
$\mhl$ {\it vs.} $X_t/\msusy$ for $\msusy=\mha=-\mu=1$~TeV for two choices
of $\tanb$ in Fig.~\ref{hhhfig4} [$\tanb=20$] and Fig.~\ref{hhhfig5}
[$\tanb=1.5$].
Note that Fig.~\ref{hhhfig4} is of particular interest, since it
allows one to read off the maximal values of $\mhl$ as a function of
$X_t$ for $\msusy\leq 1$~TeV, which were quoted in section 4.1.
The maximum value of the Higgs mass occurs for $|X_t|\simeq 2.4\msusy$.

The reader may worry that this value is too large in light of
our perturbative treatment of the squark mixing.
However, comparisons with exact diagrammatic computations confirm that
these results are trustworthy at least up to the point where the
curves reach their maxima.  From a more
practical point of view, such large values of the mixing are not very
natural; they cause tremendous splitting in the top-squark mass
eigenstates and are close to the region of parameter space where the
SU(2)$\times$U(1) breaking minimum of the
scalar potential becomes unstable relative to color and/or electromagnetic
breaking vacua \cite{casas}.

\begin{figure}[htbp]
\centerline{\psfig{file=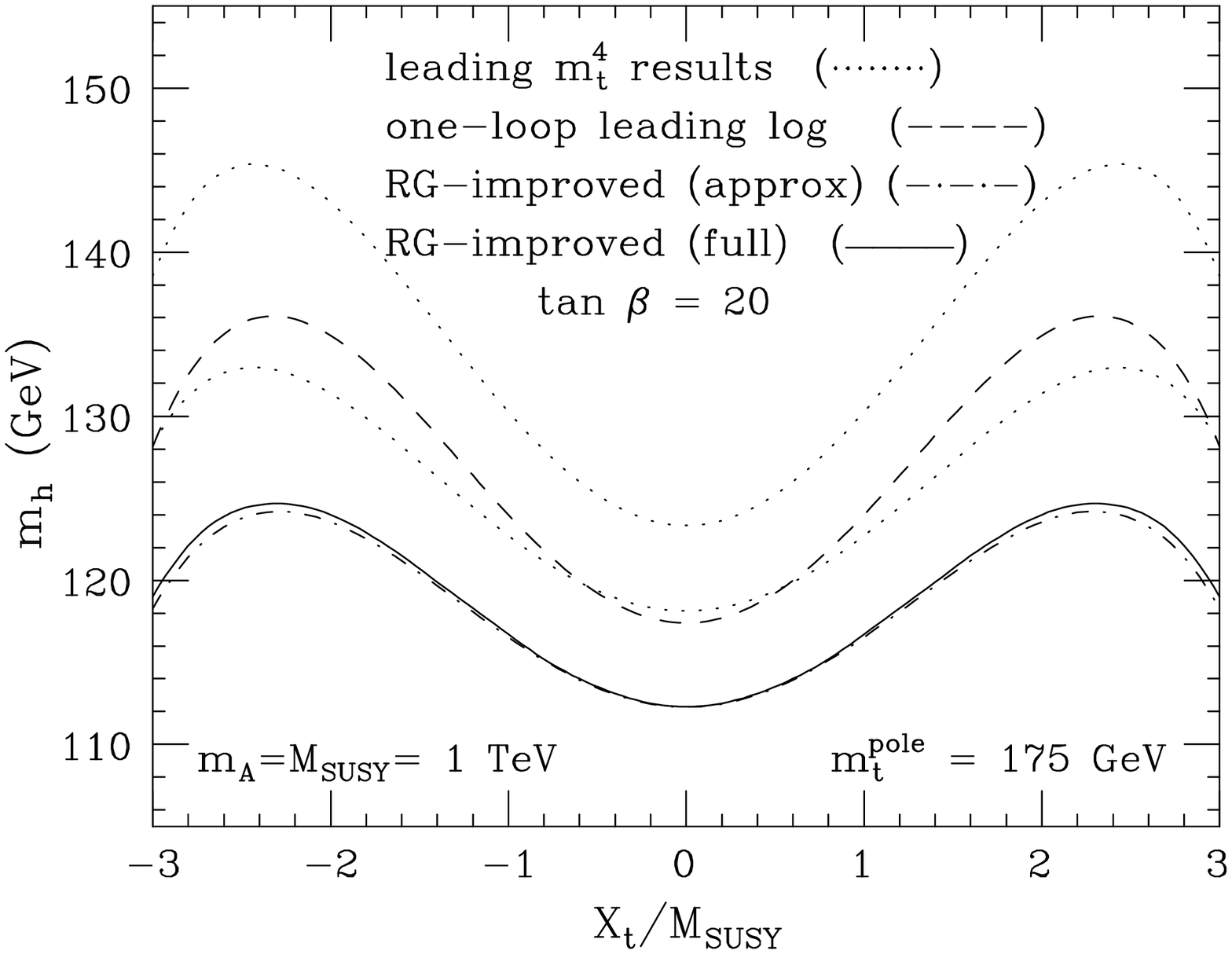,width=10cm,angle=90}}
\vskip1pc
\fcaption{The radiatively corrected light CP-even Higgs mass is plotted
as a function of $X_t/\msusy$, where $X_t\equiv A_t-\mu\cot\beta$,
for $\msusy=\mha=-\mu=1$~TeV and $\tan\beta=20$.
See the caption to Fig.~\ref{hhhfig1}.}
\label{hhhfig4}
 \vskip1pc
\centerline{\psfig{file=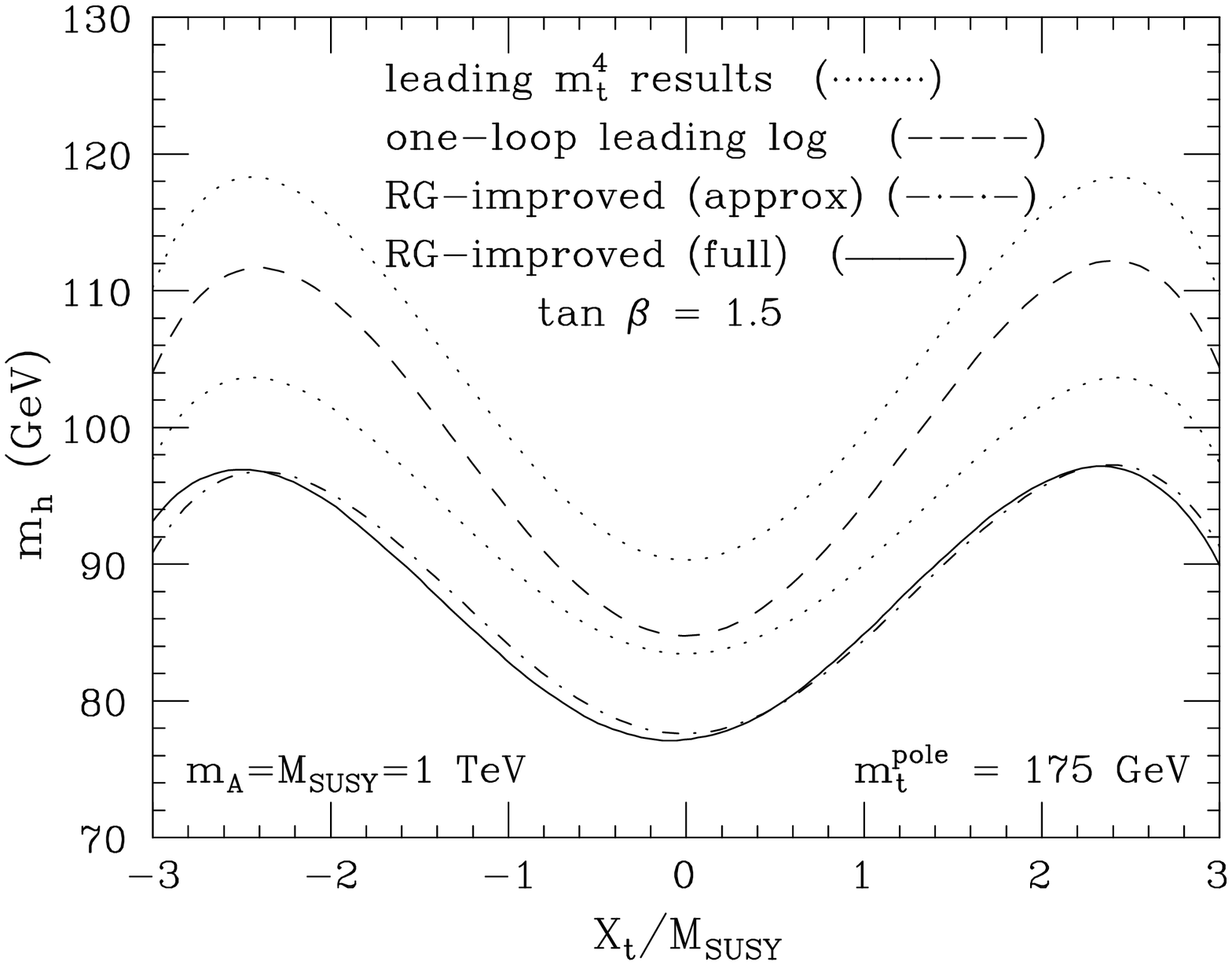,width=10cm,angle=90}}
\vskip1pc
\fcaption{The radiatively corrected light CP-even Higgs mass is plotted
as a function of $X_t/\msusy$, where $X_t\equiv A_t-\mu\cot\beta$,
for $\msusy=\mha=-\mu=1$~TeV and $\tan\beta=1.5$.
See the caption to Fig.~\ref{hhhfig1}.}
\label{hhhfig5}
\end{figure}

\begin{figure}[htbp]
\centerline{\psfig{file=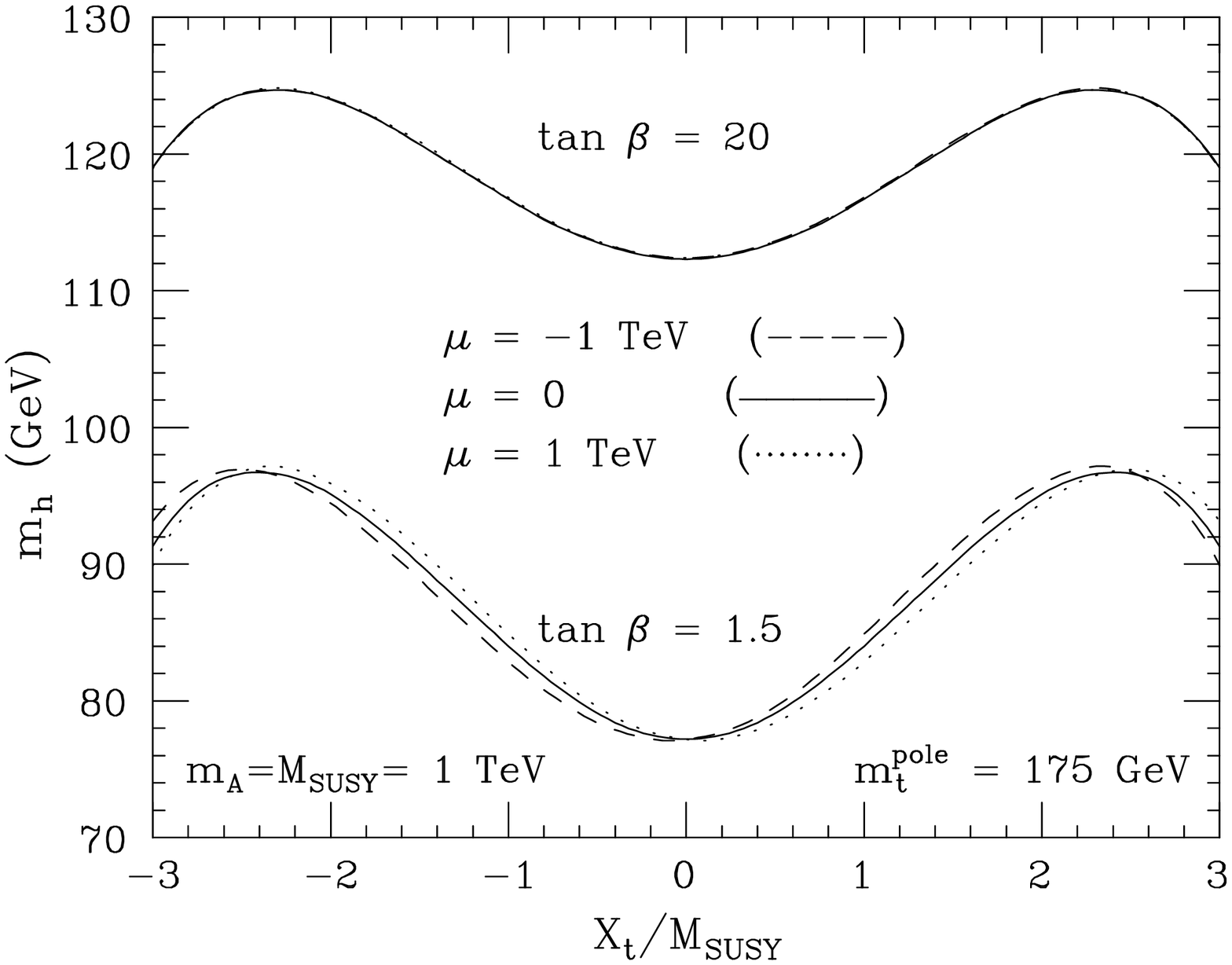,width=10cm,angle=90}}
\caption{The radiatively corrected, RG-improved
light CP-even Higgs mass is plotted
as a function of $X_t/\msusy$, where $X_t\equiv A_t-\mu\cot\beta$,
for $\msusy=\mha=1$~TeV and two choices of $\tanb=1.5$ and 20.
Three values of $\mu$ are plotted in each case: $-1$~TeV [dashed], 0
[solid] and 1~TeV [dotted].  Here, we have assumed that
the diagonal squark squared-masses are degenerate:
$M_Q=M_U=M_D=\msusy$.}
\label{hhhfig6}
 \vspace{1pc}
\centerline{\psfig{file=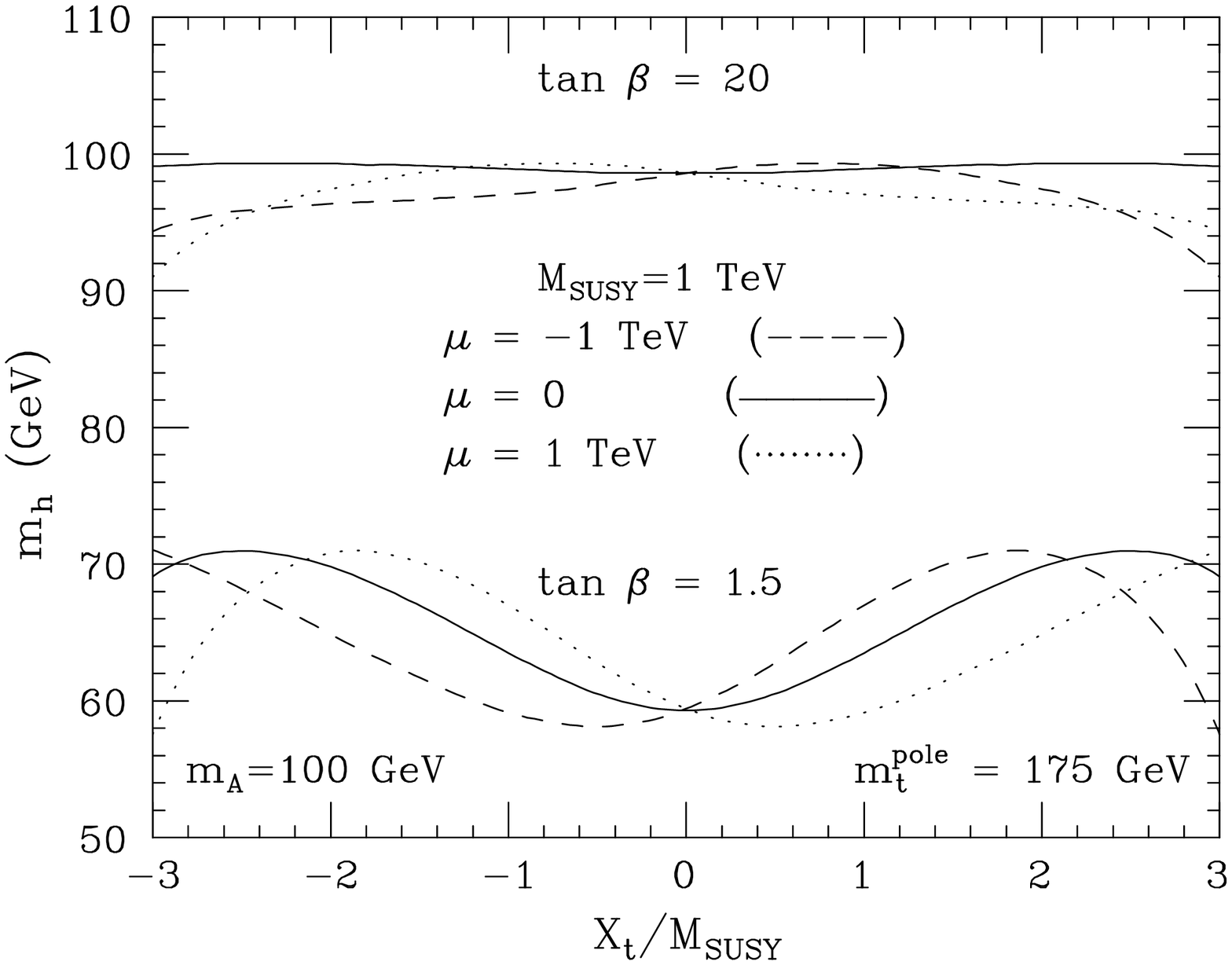,width=10cm,angle=90}}
\caption{The radiatively corrected, RG-improved
light CP-even Higgs mass is plotted
as a function of $X_t/\msusy$ for $\msusy=1$~TeV and $\mha=100$~GeV.
See the caption to Fig.~\ref{hhhfig6}.}
\label{hhhfig7}
\end{figure}

\begin{figure}[htbp]
\centerline{\psfig{file=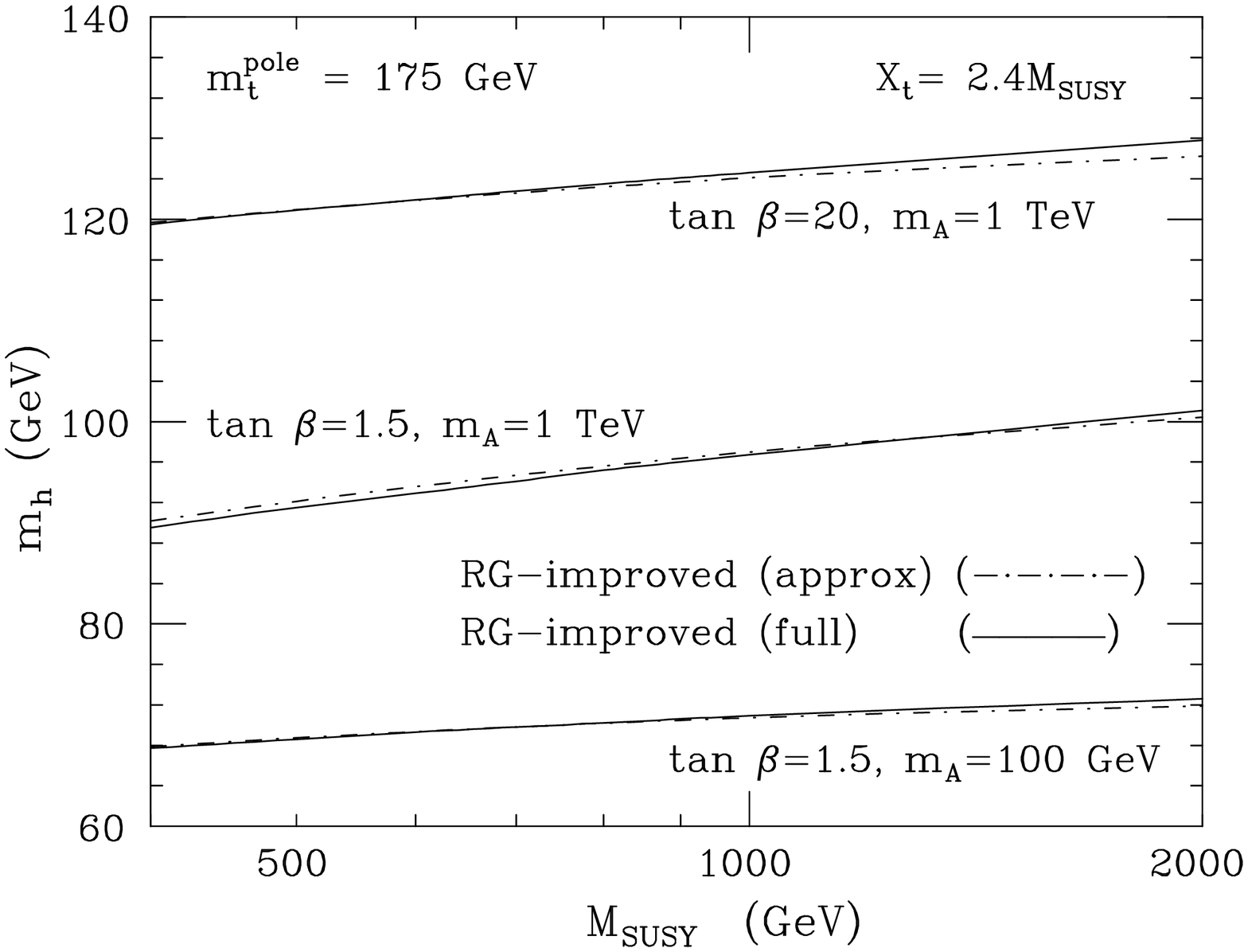,width=10cm,angle=90}}
\caption{The radiatively corrected, RG-improved
light CP-even Higgs mass is plotted
as a function of $\msusy$ for $X_t=2.4\msusy$ for three choices of
($\tanb$, $\mha$)= (20,1), (1.5,1), and (1.5,0.1), where $\mha$ is specified
in TeV units.  The solid line depicts the numerically integrated result,
and the dot-dashed line indicates the result obtained from
eq.~(\ref{simplemixform}).}
\label{hhhfig8}
 \vspace{1pc}
\centerline{\psfig{file=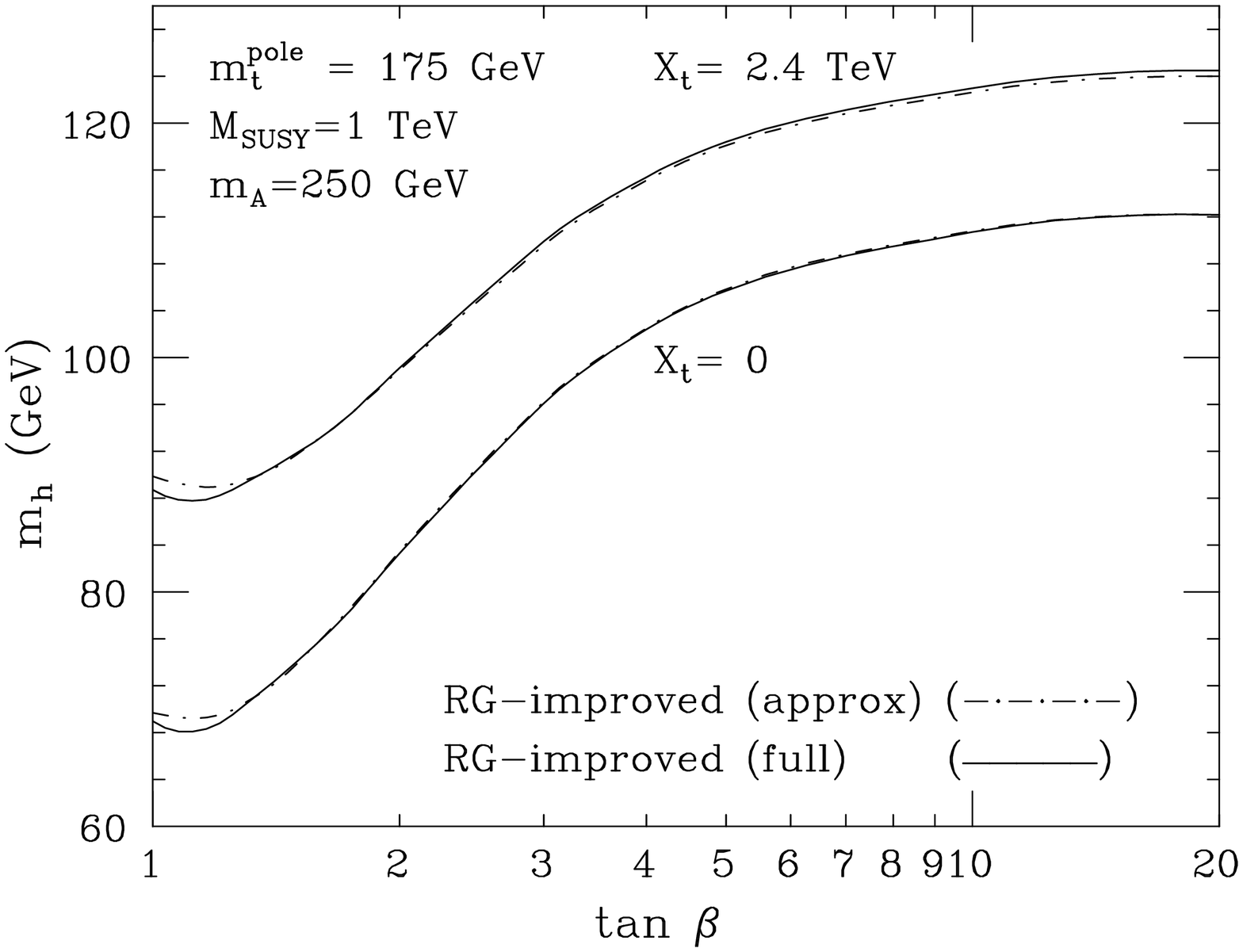,width=10cm,angle=90}}
\caption{The radiatively corrected, RG-improved
light CP-even Higgs mass is plotted
as a function of $\tanb$ for $\msusy= 1$~TeV and $\mha= 250$~GeV, for two
choices of $X_t=0$ and $X_t= 2.4\msusy$.
See the caption to Fig.~\ref{hhhfig8}.}
\label{hhhfig9}
\end{figure}

In Figs.~\ref{hhhfig4} and \ref{hhhfig5}, $\mu=-1$~TeV, {\it i.e.},
as $X_t\equiv A_t-\mu\cot\beta$ varies, so does $A_t$.
In fact, for $\mha\gg\mz$, the
dominant one-loop radiative corrections to $\mhl^2$ depend only on $X_t$
and $\msusy$ [see eq.~(\ref{deltamhs})], so that for fixed $X_t$, the
$\mu$ dependence of $\mhl$ is quite weak.  This is illustrated in
Fig.~\ref{hhhfig6}.  For values of $\mha\sim{\cal O}(\mz)$, the $\mu$
dependence is slightly more pronounced (although less so for values of
$\tanb\gg 1$) as illustrated in Fig.~\ref{hhhfig7}.
We also display $\mhl$ as a function of $\msusy$ for a number of
different parameter choices in Fig.~\ref{hhhfig8}.
In Fig.~\ref{hhhfig9}, we exhibit the $\tanb$ dependence of $\mhl$ for
two different choices of $X_t$.  Again, we notice that
our approximate formula [eq.~(\ref{simplemixform})], which is depicted by
the dot-dashed line, does remarkably well, and
never differs from the numerically integrated RG-improved value
(solid line) by more than 1.5~GeV for $\msusy\leq 2$~TeV and $\tanb\geq 1$.

In summary, when the algorithm given by eqs.~(\ref{simplemixform}) and
(\ref{scales}) is applied to the leading
log one-loop corrections plus the leading terms
resulting from squark mixing,
the full (numerically integrated) RG-improved value of
$\mhl$ is reproduced to within an accuracy of about 2~GeV (assuming that
supersymmetric particle masses lie below 2 TeV).
The methods described above also yield accurate results for the mass of
the heavier CP-even Higgs boson, $\mhh$.  The approximation to the
radiatively
corrected charged Higgs mass is slightly less accurate only because the
leading $m_t$ enhanced terms are not as dominant as in the neutral Higgs
sector.\footnote{The approximation to the radiatively
corrected charged Higgs mass
can be improved by including sub-dominant terms not contained in the
formulae given in this paper; see Ref.~\cite{madiaz} for further details.}


\section{Implications of the Radiatively Corrected Higgs Sector} 
 \label{sec:six}

Using the results of sections 4 and 5, one can obtain the leading
radiative corrections to the various Higgs couplings, and proceed
to investigate Higgs phenomenology in detail.
Here, I shall describe the procedure  used to obtain the
Higgs couplings and briefly indicate some of the consequences.
To obtain radiatively corrected couplings which are accurate in the
one-loop leading logarithmic approximation, it is sufficient to use the
tree-level couplings in which the parameters are taken to be running
parameters evaluated at the electroweak scale.  First, I remind the
reader that $\tanb$ and $\mha$ are input parameters.  Next, we obtain
the CP-even Higgs mixing angle $\alpha$ by diagonalizing the
radiatively corrected CP-even Higgs mass matrix.
With the angle $\alpha$ in hand one may compute, for example,
$\cos(\beta-\alpha)$ and $\sin\alpha$.  These results can be used
to obtain the Higgs couplings
to gauge bosons [eq.~(\ref{littletable})] and fermions [eq.~(\ref{qqcouplings})].
Finally, the Higgs self-couplings [see Appendix A] are obtained by
making use of the $\lambda_i$ evaluated at the electroweak scale.
The end result is a complete
set of Higgs boson decay widths and branching ratios that include
one-loop leading-log radiative corrections.

The Higgs production cross-section
in a two-Higgs-doublet model via the process
$e^+e^-\to Z\to Z\hh(Z\hl)$ is suppressed by a
factor $\cos^2(\beta-\alpha)$ [$\sin^2(\beta-\alpha)$]
as compared to the corresponding cross-sections in the
Standard Model.  At tree-level, we know that the decoupling limit
applies when $\mha\gg\mz$.  In fact, the approach to decoupling is quite
rapid as indicated in eq.~(\ref{largema}).
For $\mha\gsim 2\mz$, the couplings of $\hl$ to vector bosons and
to quarks and leptons are
phenomenologically indistinguishable from those of the Standard Model
Higgs boson.  Including radiative
corrections does not alter this basic behavior, although one finds that
$\cos^2(\beta-\alpha)\to 0$ more slowly as
the radiative corrections become more significant.

When radiative corrections have been incorporated, new possibilities
arise which did not exist at tree-level.  One example
is the possibility of the decay $\hl\rta\ha\ha$,
which is kinematically forbidden at tree-level but is allowed for some
range of MSSM parameters \cite{berz,nirtwo}.
We can obtain the complete one-loop leading-log
expression for the $\hl\ha\ha$ coupling (assuming $\mha \lsim m_Z$)
by inserting
the one-loop leading-log formulae for the $\lambda_i$ into
eq.~(\ref{defghaa}) \cite{nirtwo}

\vbox{%
\beqno%
&{g_{\hl\ha\ha}\over
g m_Z/2\cw}=-c_{2\beta}s_{\beta+\alpha}\left\{
1+{g^2\over96\pi^2\cw^2}\left[P_t\ln \!
\left({\msusyy\over
m_t^2}\right)+(P_b+P_f)\ln\!\left({\msusyy\over\mzz}\right)
\right]\right\}\nonumber \\[3pt]
&~~+{g^2N_c\over16\pi^2m_W^2m_Z^2}\left\{\left[
{\sa\sb^2\over\cb^3}(2m_b^4-m_b^2m_Z^2\cb^2)
-{(\ca\sb^3-\sa\cb^3)\over2\cb^2}m_b^2m_Z^2\right]
\ln\!\left({\msusyy\over m_Z^2}\right)\right.\nonumber \\[3pt]
&~~-\left.\left[{\ca\cb^2\over\sb^3}(2m_t^4-m_t^2m_Z^2\sb^2)
+{(\ca\sb^3-\sa\cb^3)\over2\sb^2}m_t^2m_Z^2\right]
\ln\!\left({\msusyy\over m_t^2}\right)\right\}\nonumber \\[3pt]
&~~-{g^2\over192\pi^2\cw^2}\left[s_{2\beta}c_{\beta+\alpha}
(P_{2H}+P_g)
-2(\ca\sb^3-\sa\cb^3)(P_{2H}'+P_g')\right]\ln\!\left({\msusyy\over
m_Z^2}\right)\,.\nonumber\\
\label{ghaall}
\eeqno
}

\noindent If kinematically allowed, $\hl\rta\ha\ha$ would almost
certainly be the dominant decay mode.  However, the LEP experimental
lower
bound on $\mha$ now lies above $0.5(\mhl)_{\rm max}\simeq 62.5$~GeV.
Thus, the region of parameter space  where the decay $\hl\to\ha\ha$ is
kinematically allowed is no longer viable.  The possibility of
measuring the
$\hl\ha\ha$ couplings at a future $e^+e^-$ linear collider by detecting
double Higgs production has been discussed in Ref.~\cite{djouadi}.
Unfortunately, the prospects are poor due to low cross-sections and
significant backgrounds.

For the heavier Higgs states, there are many possible
final state decay modes.  The various branching ratios are complicated
functions of the MSSM parameter space \cite{gbhs}.  For example, a plot
of the branching ratios of $\hh$, with the leading one-loop
radiative corrections included, can be found in Ref.~\cite{gsw}.
This plot
indicates a rich phenomenology for heavy Higgs searches at
future colliders.  The precision measurements of Higgs masses
and couplings will be one of the primary tasks of the LHC and future
lepton-lepton colliders \cite{snowmass,gunreport}

Although the
possibility of a light Higgs discovery at LEP still remains, the effects
of the radiative corrections may be significant enough
to push the Higgs boson above the LEP-2 discovery reach.
In this case, the discovery of the Higgs boson will be the purview of
the LHC. Of course, if low-energy
supersymmetry exists, then LHC will also uncover
direct evidence for the supersymmetric particles.
In this case, a detailed examination of the Higgs sector, with precision
measurements of the Higgs masses and couplings, will provide a critical
test for the underlying supersymmetric structure.  Unlocking the secrets
of the Higgs bosons will help reveal
the mechanism of electroweak symmetry breaking and the nature
of the TeV scale physics that lies beyond the Standard Model.

\vskip3pc
\centerline{{\bf Acknowledgments}}
\medskip
I would like to express my deep appreciation to
my collaborators Ralf Hempfling and Andre Hoang,
whose contributions to the work on radiatively corrected
Higgs masses were instrumental to the development of the material
reported in this paper.  I would also like to thank
Marco D\'\i az, Abdel Djouadi, Yuval Grossman, Jack Gunion, Yossi Nir,
Scott Thomas, and Peter Zerwas for many fruitful interactions.
Finally, I gratefully acknowledge conversations with Marcela Carena,
Mariano Quiros and Carlos Wagner and appreciate the opportunity
provided by the
1995 LEP-2 workshop to revisit many of the issues discussed here.
This work was supported in part by the U.S. Department of Energy.
\clearpage
\makeatletter
\@addtoreset{equation}{section}
\def\theequation{\thesection.\arabic{equation}}
\makeatother

            \appendix 
\centerline{\bf Appendix}
\vspace{-1pc}
\section{Three-Higgs Vertices in the Two-Higgs Doublet Model}
      \label{app:A}

In this Appendix, I list the Feynman rules for the 3-point Higgs
interaction in the most general (nonsupersymmetric)
two-Higgs doublet extension of the Standard Model, assuming that
the Higgs sector conserves CP.
The Feynman rule for the $ABC$ vertex is denoted by $ig\ls{ABC}$.
For completeness, $R$-gauge Feynman rules
involving the Goldstone bosons ($G^\pm$ and $G^0$)
are also listed.

\vskip-2pt
Interactions involving physical Higgs bosons depend in detail on the
parameters of the Higgs potential specified in eq.~(\ref{pot}).
\beqno%
g\ls{h^0A^0A^0} &=
   {2\mw\over g}\bigl[ \lambda_1\sb^2\cb\sa
   -\lambda_2\cb^2\sb\ca -\widetilde\lambda_3
   (\sb^3\ca-\cb^3\sa) +2\lambda_5\sba \nonumber\\
&\qquad -\lambda_6\sb\big(\cb\sab+\sa
c_{2\beta}\big)-\lambda_7\cb\big(\ca c_{2\beta}-\sb\sab\big)\bigr]\,,\nonumber \\[3pt]
g\ls{H^0A^0A^0} &= {-2\mw\over g}\bigl[
   \lambda_1\sb^2\cb\ca+\lambda_2\cb^2\sb\sa+\widetilde\lambda_3
   (\sb^3\sa+\cb^3\ca) -2\lambda_5\cba \nonumber\\
&\qquad -\lambda_6\sb\big(\cb\cab+\ca
c_{2\beta}\big)+\lambda_7\cb\big(\sb\cab+\sa c_{2\beta}\big)\bigr]\,,\nonumber \\[3pt]
g\ls{h^0H^0H^0} &= {6\mw\over g}\bigl[
   \lambda_1\ca^2\cb\sa-\lambda_2\sa^2\sb\ca+\widetilde\lambda_3
   (\sa^3\cb-\ca^3\sb+\twothirds\sba) \nonumber\\
&\quad -\lambda_6\ca\big(\cb c_{2\alpha}-\sa\sab\big)
-\lambda_7\ca\big(\sb c_{2\alpha}+\ca\sab\big)\bigr]\,,\nonumber \\[3pt]
g\ls{H^0h^0h^0} &= {-6\mw\over g}\bigl[
   \lambda_1\sa^2\cb\ca+\lambda_2\ca^2\sb\sa+\widetilde\lambda_3
   (\sa^3\sb+\ca^3\cb-\twothirds\cba) \nonumber\\
&\quad -\lambda_6\sa\big(\cb c_{2\alpha}+\ca\cab\big)
+\lambda_7\ca\big(\sb
c_{2\alpha}+\sa\cab\big)\bigr]\,,\nonumber\\[3pt]
g\ls{h^0h^0h^0} &= {6\mw\over g}\bigl[
   \lambda_1\sa^3\cb-\lambda_2\ca^3\sb+\widetilde\lambda_3
   \sa\ca\cab \nonumber\\
&\quad -\lambda_6\sa^2\big(3\ca\cb-\sa\sb\big)
+\lambda_7\ca^2\big(3\sa\sb-\ca\cb\big)\bigr]\,,\nonumber \\[3pt]
g\ls{H^0H^0H^0} &= {-6\mw\over g}\bigl[
   \lambda_1\ca^3\cb+\lambda_2\sa^3\sb+\widetilde\lambda_3
   \sa\ca\sab \nonumber\\
&\quad +\lambda_6\ca^2\big(3\sa\cb+\ca\sb\big)
+\lambda_7\sa^2\big(3\ca\sb+\sa\cb\big)\bigr]\,,\nonumber \\[3pt]
g\ls{h^0H^+H^-} &=g\ls{h^0A^0A^0}-{2\mw\over g}
   \big(\lambda_5-\lambda_4\big)\sba\,,\nonumber \\[3pt]
g\ls{H^0H^+H^-} &=g\ls{H^0A^0A^0}-{2\mw\over g}
   \big(\lambda_5-\lambda_4\big)\cba\,,
\label{defghaa}
\eeqno

\noindent
where I have used the notation
\beq
\widetilde\lambda_3\equiv\lambda_3+\lambda_4+\lambda_5\,.
\label{lthree}
\eeq
It is interesting to note that couplings of the charged Higgs bosons
satisfy relations analogous to that of $\mhpm$ given in eq.~(\ref{mamthree}).

The Feynman rules for three-point Higgs vertices that involve Goldstone
bosons exhibit much simpler forms
\beqno
g\ls{h^0G^0G^0} &=
   {-g\over 2\mw} \mhl^2\sin(\beta-\alpha)\,,\nonumber \\[3pt]
g\ls{H^0G^0G^0} &=
   {-g\over 2\mw} \mhh^2\cos(\beta-\alpha)\,,\nonumber \\[3pt]
g\ls{h^0G^+G^-} &=g\ls{h^0G^0G^0}\,,\nonumber \\[3pt]
g\ls{H^0G^+G^-} &=g\ls{H^0G^0G^0}\,,\nonumber \\[3pt]
g\ls{h^0A^0G^0} &= {-g\over 2\mw}(\mhl^2-\mha^2)\cos(\beta-\alpha)\,,
\nonumber \\[3pt]
g\ls{H^0A^0G^0} &= {g\over 2\mw}(\mhh^2-\mha^2)\sin(\beta-\alpha)\,,\nonumber \\[3pt]
g\ls{h^0H^\pm G^\mp} &=
{g\over 2\mw}(\mhpm^2-\mhl^2)\cos(\beta-\alpha)\,,\nonumber \\[3pt]
g\ls{H^0H^\pm G^\mp} &=
{-g\over 2\mw}(\mhpm^2-\mhh^2)\sin(\beta-\alpha)\,,\nonumber \\[3pt]
g\ls{A^0H^\pm G^\mp} &=
{\pm g\over 2\mw}(\mhpm^2-\mha^2)\,.
\label{goldstonerules}
\eeqno
  \vskip6pt\noindent
In the rule for the $\ha H^\pm G^\mp$ vertex, the sign corresponds to
$H^\pm$ entering the vertex and $G^\pm$ leaving the vertex.

One can easily check that if tree-level MSSM relations are imposed
on the $\lambda_i$, Higgs masses, and angles $\alpha$ and $\beta$,
one recovers the MSSM Feynman rules listed in Appendix A of
Ref.~\cite{hhg}.
\clearpage
\section{Renormalization Group Equations}  \label{app:B}

In this Appendix, I have collected
the one-loop renormalization group equations (RGEs)
that are needed in the analysis presented in this paper
 \cite{chengli,rgecollection,llog}.
Schematically, the RGEs at one-loop take the form
\beq
{dp_i\over dt}=\beta_i(p_1,p_2,..)\,,\qquad
\mbox{where}~t\equiv\ln\,\mu^2\,,\label{rges}
\eeq
where $\mu$ is the energy scale, and
the parameters $p_i$ stand for
the Higgs boson self-couplings $\lambda_i$ ($i=1\ldots 7$),
the squared Yukawa couplings $h_f^2$
($f=t$, $b$ and $\tau$; the two lighter generations can be neglected),
and the squared gauge couplings $g_j^2$ ($j=$3, 2, 1) corresponding to
SU(3)$\times$SU(2)$\times$U(1) respectively.  The $g_j$ are
normalized such that they are equal at the grand unification
scale.  It is also convenient to define
\beq
g\equiv g\ls2\,,\qquad\qquad g^\pri\equiv \sqrt{\threefifths} g\ls1\,,
\label{gaugecoups}
\eeq
where $g$ and $g^\pri$ are normalized in the usual way for low-energy
electroweak physics, {\it i.e.} $\tan\theta_W=g^\pri/g$.

I now list the $\beta$-functions required for the analysis presented
in this paper.  Two cases will be given, depending on whether $\mu$ is
above or below the scale of supersymmetry breaking, $\msusy$.\\
\bigskip
1. $\mu>\msusy$
\beqno%
\beta_{h_t^2}&={h_t^2\over16\pi^2}\left[6
h_t^2+ h_b^2-\sixteenthirds g_3^2-3g^2-\thirteennineths g'^2\right]
\nonumber \\[3pt]
\beta_{h_b^2}&={h_b^2\over16\pi^2}\left[6 h_b^2+h_t^2
+h_{\tau}^2-\sixteenthirds g_3^2-3g^2-\sevennineths g'^2\right]
\nonumber \\[3pt]
\beta_{h_{\tau}^2}&={h_{\tau}^2\over16\pi^2}\left[4
h_{\tau}^2+3 h_b^2-3g^2-3g'^2\right]\nonumber\\[3pt]
\beta_{g'^2}&={g^{\pri4}\over48\pi^2}\Big[10N_g+\threehalf N_H\Big]
\nonumber \\[3pt]
\beta_{g^2}&={g^4\over48\pi^2}\Big[6N_g+\threehalf N_H-18\Big]
\nonumber \\[3pt]
\beta_{g_3^2}&={g_3^4\over48\pi^2}\Big[6N_g-27\Big]\,.
\label{betagg}
\eeqno
Here $N_g=3$ is the number of generations,
$N_H=2$ is the number of scalar doublets,
and the Higgs-fermion Yukawa couplings are given by

\vbox{%
\beqno%
h_t&={gm_t\over \sqrt2 m_W \sinb}\,,  \nonumber \\[3pt]
h_{d_i}&={gm_{d_i}\over \sqrt2 m_W \cosb}\,,\qquad (d_i =
b,\tau)\,.
\label{yukawas}
\eeqno
}

\bigskip
2. $\mu<\msusy$
\beqno%
\beta_{h_t^2}      &={h_t^2\over16\pi^2}\big[\ninehalf
h_t^2+\half h_b^2-8g_3^2-\ninefourth g^2-\seventeentwelfth g'^2\big]
\nonumber \\[3pt]
\beta_{h_b^2}      &={h_b^2\over 16\pi^2}\big[\ninehalf h_b^2+\half
h_t^2+h_{\tau}^2-8g_3^2-\ninefourth g^2-\fivetwelfth g'^2\big]
\nonumber \\[3pt]
\beta_{h_{\tau}^2} &={h_{\tau}^2\over16\pi^2}\big[\fivehalf
h_{\tau}^2+3h_b^2-\ninefourth g^2 -\fifteenfourth g'^2\big]
\nonumber \\[3pt]
\beta_{g'^2}  &={g^{\pri4}\over48\pi^2}\Big[\twentythirds N_g+\half
N_H\Big]  \nonumber \\[3pt]
\beta_{g^2}   &={g^4\over48\pi^2}\Big[4N_g+\half N_H-22\Big]
\nonumber \\[3pt]
\beta_{g_3^2} &={g_3^4\over48\pi^2}\Big[4N_g-33\Big]\,.
\label{betagg2}
\eeqno
In writing down the
RGEs for the Higgs-fermion Yukawa couplings in eq.~(\ref{betagg2}), I
have assumed
that the Higgs-fermion interaction is the same as in the MSSM; namely,
$\Phi_1$ [$\Phi_2$] couples exclusively to down-type
[up-type] fermions.  Moreover, in deriving the $\mu<\msusy$ equations,
it was assumed that
the effective low-energy theory at the scale $\mu$ includes the full
two-doublet Higgs sector (but does not include the supersymmetric
particles, whose masses are of order $\msusy$).

Finally, I list the RGEs for the
Higgs self-couplings of the general two-Higgs doublet model (with
the Higgs-fermion couplings as specified above).
First, I need to define the anomalous dimensions of the two Higgs
fields:
\beqno%
\gamma_{1} &={1\over 64\pi^2}
   \Big[9g^2+3g'^2-4\sum_i N_{ci}h_{d_i}^2\Big]\,,\nonumber \\[3pt]
\gamma_{2} &={1\over 64\pi^2}
    \Big[9g^2+3g'^2-4\sum_i N_{ci}h_{u_i}^2\Big]\,,
\label{wavez}
\eeqno
where the sum over $i$ is taken over three generations of quarks
(with $N_c=3$) and leptons (with $N_c=1$).  The $\beta$-functions
for the Higgs self-couplings in the general CP-conserving
non-supersymmetric two-Higgs-doublet model (with
the Higgs-fermion couplings as specified in section 2)
are given by

\vbox{%
\beqno%
\beta_{\lambda_{1}} &={1\over16\pi^2}
   \bigg\{ 6\lambda^2_{1}+2\lambda_3^2+2\lambda_3\lambda_4+
   \lambda_4^2+\lambda_5^2+12\lambda_{6}^2 \bigg.\nonumber \\[3pt]
&\hskip2cm \bigg.        +\threeighth
       \big[2g^4+(g^2+g'^2)^2\big]-2\sum_i N_{ci}h_{d_i}^4\bigg\}
      -2\lambda_{1}\gamma_{1}\nonumber \\[3pt]
\beta_{\lambda_{2}} &={1\over16\pi^2}
   \bigg\{ 6\lambda^2_{2}+2\lambda_3^2+2\lambda_3\lambda_4+
   \lambda_4^2+\lambda_5^2+12\lambda_{7}^2 \bigg.\nonumber \\[3pt]
&\hskip2cm \bigg. +\threeighth
      \big[2g^4+(g^2+g'^2)^2\big]-2\sum_i N_{ci}h_{u_i}^4\bigg\}
      -2\lambda_{2}\gamma_{2}\nonumber \\[3pt]
\beta_{\lambda_3} &={1\over16\pi^2}
   \bigg\{ (\lambda_1+\lambda_2)(3\lambda_3+\lambda_4)+2\lambda_3^2
   +\lambda_4^2+\lambda_5^2+2\lambda_6^2+2\lambda_7^2
   +8\lambda_6\lambda_7 \bigg.\nonumber \\[3pt]
&\hskip2cm +\threeighth   \bigg.
      \big[2g^4+(g^2-g'^2)^2\big]-2\sum_i N_{ci}h_{u_i}^2h_{d_i}^2\bigg\}
      -\lambda_3(\gamma_1+\gamma_2)\nonumber \\[3pt]
\beta_{\lambda_4}  &={1\over16\pi^2}
   \Big[ \lambda_4(\lambda_1+\lambda_2+4\lambda_3+2\lambda_4)+
   4\lambda_5^2+5\lambda_6^2+5\lambda_7^2+2\lambda_6\lambda_7\Big.\nonumber \\[3pt]
&\hskip2cm +\threehalf       \Big.
      g^2g'^2+2\sum_i N_{ci}h_{u_i}^2h_{d_i}^2\Big]
      -\lambda_4(\gamma_1+\gamma_2)\nonumber \\[3pt]
\beta_{\lambda_5} &={1\over16\pi^2}
   \Big[\lambda_5(\lambda_1+\lambda_2+4\lambda_3+6\lambda_4)
   +5\big(\lambda_6^2+\lambda_7^2\big)+2\lambda_6\lambda_7\Big]
   -\lambda_5(\gamma_1+\gamma_2)\nonumber \\[3pt]
\beta_{\lambda_6} &={1\over16\pi^2}
   \Big[\lambda_6(6\lambda_1+3\lambda_3+4\lambda_4+5\lambda_5\big)
   +\lambda_7\big(3\lambda_3+2\lambda_4+\lambda_5\big)\Big]
   -\half\lambda_6(3\gamma_1+\gamma_2)\nonumber \\[3pt]
\beta_{\lambda_7} &={1\over16\pi^2}
   \Big[\lambda_7(6\lambda_2+3\lambda_3+4\lambda_4+5\lambda_5\big)
   +\lambda_6\big(3\lambda_3+2\lambda_4+\lambda_5\big)\Big]
   -\half\lambda_7(\gamma_1+3\gamma_2)\,.   \nonumber\\
\label{betal}
\eeqno
}

\clearpage



\begin{thebibliography}{000}


\bibitem{hhg}
J.F. Gunion, H.E. Haber, G. Kane and S. Dawson,
{\it The Higgs Hunter's Guide} (Addison-Wesley Publishing Company,
Reading, MA, 1990).
\bibitem{habertasi}
H.E. Haber, in {\it Testing the Standard Model,}
Proceedings of the 1990 Theoretical Advanced Study Institute
in Elementary Particle Physics, edited by M. Cveti\v c  and
P. Langacker (World Scientific, Singapore, 1991) p. 340--475.
\bibitem{thooft}
G. 't Hooft, in {\it Recent Developments in Gauge
Theories,} Proceedings of the NATO Advanced Summer Institute,
Cargese, 1979, edited by G. 't~Hooft \etal\ (Plenum, New York,
1980) p.~135--157.
\bibitem{suss}
L. Susskind, {\sl Phys. Rep.} {\bf 104} (1984) 181.
\bibitem{susysol}
E.~Witten, \sl Nucl.~Phys. \bf B188 \rm (1981) 513;
S.~Dimopoulos and H.~Georgi, \sl Nucl.~Phys. \bf B193 \rm (1981) 150;
N.~Sakai, \sl Z.~Phys. \bf C11 \rm (1981) 153; R.K. Kaul,
{\sl Phys. Lett.} {\bf 109B} (1982) 19.
\bibitem{susyrev}
H.P. Nilles, {\sl Phys.~Rep.} {\bf 110} (1984) 1;
H.E. Haber and G.L. Kane, {\sl Phys.~Rep.} {\bf 117} (1985) 75;
A.B. Lahanas and D.V. Nanopoulos, {\sl Phys. Rep.} {\bf 145} (1987) 1;
R. Barbieri, {\sl Riv. Nuovo Cimento} {\bf 11} (1988) 1;
R. Arnowitt and P. Nath, in {\sl
Particles and Fields}, Proceedings of the VII Jorge Andre
Swieca Summer School, Sao Paulo, Brazil, 10--23 January, 1993, edited by
J.P. Eboli and V.O. Rivelles (World Scientific, Singapore, 1994)
pp.~3--63.
\bibitem{hehtasi}
H.E. Haber, \prd{54}{1996}{687};
H.E. Haber, in {\it Recent Directions in Particle Theory}, Proceedings
of the 1992 Theoretical Advanced Study Institute in Elementary Particle
Physics, edited by J. Harvey and J. Polchinski (World Scientific,
Singapore, 1993) pp.~589--686.
\bibitem{hhgref}
For a comprehensive review and a guide to the literature,
see Chapter 4 of Ref.~\cite{hhg}.
\bibitem{ghw}%
J.M. Cornwall, D.N. Levin and G. Tiktopoulos, \NPRL 30&73&1268&;
\NPR D10&74&1145&;
C.H. Llewellyn Smith, \NPL 46B&73&233&;
H.A. Weldon, \NPR D30&84&1547&;
J.F. Gunion, H.E. Haber and J. Wudka, {\sl Phys. Rev.}
{\bf D43} (1991) 904.
\bibitem{hhgsusy}
J.F. Gunion and H.E. Haber, {\sl Nucl. Phys.} {\bf B272}
(1986) 1; {\bf B278} (1986) 449 [E: {\bf B402} (1993) 567].
\bibitem{gw}
S. Glashow and S. Weinberg, {\sl Phys. Rev.} {\bf D15}
(1977) 1958; E.A. Paschos, {\sl Phys. Rev.} {\bf D15}
(1977) 1966.
\bibitem{habernir} {H. E. Haber and Y. Nir},
{\sl Nucl. Phys.} {\bf B335} (1990) 363.
%
\bibitem{DECP}
{H.E. Haber}, in {\it Beyond the Standard Model IV},
Proceedings of the Fourth International Conference
on Physics Beyond the Standard Model, Granlibakken,
Lake Tahoe, CA, 13--18 December, 1994, edited by J.F. Gunion,
T. Han and J. Ohnemus (World Scientific, Singapore, 1995) pp.~151--163;
and in {\it Perspectives for Electroweak Interactions in $e^+e^-$
Collisions}, Proceedings of the Ringberg Workshop, Ringberg Castle,
Tegernsee, Germany, 5--8 February, 1995, edited by B.A. Kniehl
(World Scientific, Singapore, 1995) pp.~219--231.
%
\bibitem{LEPHIGGS}
J.P. Martin, LYCEN-9644 (1996), in Proceedings of the 28th International
Conference on High Energy Physics, Warsaw, Poland, 25--31 July 1996,
edited by Z. Ajduk and A.K. Wroblewski (World Scientific, Singapore, 1997).
%
\bibitem{janot}
M. Carena, P.M. Zerwas {\it et al.}, in {\it Physics at LEP2},
Volume 1, edited by G. Altarelli, T. Sj\"ostrand and F. Zwirner,
CERN Yellow Report 96-01 (1996) pp.~351--462.
%
%
\bibitem{ypan}
R.~Barate {\it et al.} [ALEPH Collaboration], CERN-PPE/97-70 (1997);
CERN-PPE/97-71 (1997), submitted to {\sl Phys. Lett. B}.
%
\bibitem{grant}
A.K. Grant, {\sl Phys. Rev.} {\bf D51} (1995) 207.
%
\bibitem{joanne}
J.L. Hewett, {\sl Phys. Rev. Lett.} {\bf 70} (1993) 1045;
V. Barger, M. Berger and R.J.N. Phillips, {\sl Phys. Rev. Lett.} {\bf
70} (1993) 1368; J.L. Hewett, in {\it B Physics: Physics Beyond
the Standard Model at the B Factory}, Proceedings of the International
Workshop on $B$ Physics,
Nagoya, Japan, 26--28 October 1994, edited by
A.I. Sanda and S. Suzuki (World Scientific, Singapore, 1995)
pp.~321--326.
%
\bibitem{ghn}
Y. Grossman, H.E. Haber and Y. Nir, {\sl Phys. Lett.} {\bf B357}
(1995) 630.
%
\bibitem{sola}
J.A. Coarasa, R.A. Jim\'enez and J. Sol\`a, UAB-FT-407 (1997)
[hep-ph/9701392].
%
\bibitem{hewett}
V. Barger, J.L. Hewett and R.J.N. Phillips, {\sl Phys.
Rev.} {\bf D41} (1990) 3421.
%
\bibitem{schrempp}
B. Schrempp and M. Wimmer, {\sl Prog. Part.
Nucl. Phys.} {\bf 37} (1996) 1.
%
\bibitem{hhprl}
H.E. Haber and R. Hempfling, {\sl Phys. Rev. Lett.} {\bf 66}
(1991) 1815.
\bibitem{early-veff}
 Y. Okada, M. Yamaguchi and T. Yanagida, \ptp{85}{1991}{1};
 J. Ellis, G. Ridolfi and F. Zwirner, \plb{257}{1991}{83}.

\bibitem{veff}
S.P. Li and M. Sher, \plb{140}{1984}{339};
R. Barbieri and M. Frigeni, {\sl Phys. Lett.} {\bf B258} (1991) 395;
 M. Drees and M.M. Nojiri, \prd{45}{1992}{2482};
 J.A. Casas, J.R. Espinosa, M. Quiros and A. Riotto,
 \npb{436}{1995}{3} [E: {\bf B439} (1995) 466].
%
\bibitem{berz}
A. Brignole, J. Ellis, G. Ridolfi and F. Zwirner,
{\sl Phys. Lett.} {\bf B271} (1991) 123; [E: {\bf B273} (1991) 550].
%
\bibitem{erz}
J. Ellis, G. Ridolfi and F. Zwirner, \plb{262}{1991}{477}.


\bibitem{carena}
M. Carena, J.R. Espinosa, M. Quiros and C.E.M. Wagner,
{\sl Phys. Lett.} {\bf B355} (1995) 209;  M. Carena, M. Quiros and
C.E.M. Wagner, {\sl Nucl. Phys.} {\bf B461} (1996) 407.

\bibitem{turski}
J.F. Gunion and A. Turski, \prd{39}{1989}{2701}; \prd{40}{1989}{2333}.

%
\bibitem{brig}
A. Brignole, \plb{277}{1992}{313}.
%
\bibitem{madiaz}
M.A. D\'\i az and H.E. Haber, \prd{45}{1992}{4246}.
%
\bibitem{1-loop}
M.S. Berger, \prd{41}{1990}{225}; A. Brignole, \plb{281}{1992}{284};
M.A. D\'\i az and H.E. Haber, \prd{46}{1992}{3086}.

\bibitem{hempfhoang}
 R. Hempfling and A.H. Hoang, \plb{331}{1994}{99}.


%
\bibitem{completeoneloop}
 P.H. Chankowski, S. Pokorski and J. Rosiek, \plb{274}{1992}{191};
  \npb{423}{1994}{437};
  A. Yamada, \plb{263}{1991}{233}; \zpc{61}{1994}{247};
 A. Dabelstein, \zpc{67}{1995}{495};
D.M. Pierce, J.A. Bagger, K. Matchev and R. Zhang,
{\sl Nucl. Phys.} {\bf B491} (1997) 3.

%

\bibitem{rge}
R. Barbieri, M. Frigeni and F. Caravaglios, \plb{258}{1991}{167};
 Y. Okada, M. Yamaguchi and T. Yanagida, \plb{262}{1991}{54};
 D.M. Pierce, A. Papadopoulos and S. Johnson, \prl{68}{1992}{3678};
 K. Sasaki, M. Carena and C.E.M. Wagner, \npb{381}{1992}{66};
 R. Hempfling, in {\it Phenomenological Aspects of Supersymmetry},
 edited by W. Hollik, R. R\"uckl and J. Wess (Springer-Verlag, Berlin,
 1992) p.~260--279;
 J. Kodaira, Y. Yasui and K. Sasaki, \prd{50}{1994}{7035}.
%
\bibitem{2loopquiros}
J.R. Espinosa and M. Quiros, \plb{266}{1991}{389}.
%
\bibitem{llog}H.E. Haber and R. Hempfling, \prd{48}{1993}{4280}.
%
\bibitem{hhh}
H.E. Haber, R. Hempfling and A.H. Hoang, hep/ph-9609331 (1996), 
{\sl Z. Phys. C} (1997), in press.
%
\bibitem{chengli}
T.P. Cheng, E. Eichten and L.-F. Li, {\sl Phys. Rev.}
{\bf D9} (1974) 2259.
\bibitem{cmpp}
N. Cabibbo, L. Maiani, G. Parisi and R. Petronzio
\sl Nucl. Phys. \bf B158 \rm (1979) 295.
%
\bibitem{casas}
For a recent study of color and electric charge breaking minima in the
MSSM and references to earlier works, see J.A. Casas, A. Lleyda and
C. Mu\~noz, \npb{471}{1996}{3}.
%
\bibitem{nirtwo}
H.E. Haber, R. Hempfling and Y. Nir, {\sl Phys. Rev.} {\bf D46} (1992)
3015.
%
\bibitem{djouadi}
A. Djouadi, H.E. Haber and P.M. Zerwas, {\sl Phys. Lett.}
{\bf B375} (1996) 203; F. Boudjema and E. Chopin, {\sl Z. Phys.} {\bf
C73} (1996) 85.
%
\bibitem{gbhs}
V. Barger, M.S. Berger, A.L. Stange and R.J.N. Phillips,
{\sl Phys. Rev.} {\bf D45} (1991) 4128;
J.F. Gunion, R. Bork, H.E. Haber and A. Seiden,
{\sl Phys. Rev.} {\bf D46} (1992) 2040;
J.F. Gunion, H.E. Haber and C. Kao,
{\sl Phys. Rev.} {\bf D46} (1992) 2907;
Z. Kunszt and F. Zwirner, {\sl Nucl.Phys.} {\bf B385} (1992) 3;
A. Yamada, {\sl Mod. Phys. Lett.} {\bf A7} (1992) 2877.
%
\bibitem{gsw}
J.F. Gunion, A. Stange and S. Willenbrock, 
in {\it Electroweak Symmetry Breaking and
New Physics at the TeV Scale}, edited
by T.L. Barklow, S. Dawson, H.E. Haber, and J.L. Siegrist
(World Scientific, Singapore, 1996) pp.~23--145.
%
\bibitem{snowmass}
H.E. Haber {\it et al.}, hep-ph/9703391 (1997),
summary report of the Weakly-Coupled Higgs
Boson and Precision Electroweak Physics Working Group, to appear in the
Proceedings of the 1996 Snowmass Workshop on New Directions in High
Energy Physics, 25 June--12 July, 1996.
%
\bibitem{gunreport}
J.F. Gunion {\it et al.}, hep-ph/9703330 (1997),
Higgs Boson Discovery and Properties Subgroup
Report, to appear in the Proceedings of the 1996 Snowmass Workshop, {\it
op.~cit.}
%
\bibitem{rgecollection}
K. Inoue, A. Kakuto and Y. Nakano,
{\sl Prog. Theor. Phys.} {\bf 63} (1980) 234; H. Komatsu,
{\sl Prog. Theor. Phys.} {\bf 67} (1982) 1177; K. Inoue, A. Kakuto,
H. Komatsu and S. Takeshita,
{\sl Prog. Theor. Phys.} {\bf 67} (1982) 1889; {\bf 68} (1982) 927
[E: {\bf70} (1983) 330] {\bf 71} (1984) 413; C. Hill, C.N. Leung and
S. Rao, {\sl Nucl. Phys.} {\bf B262} (1985) 517;
J. Bagger, S. Dimopoulos and  E. Masso, {\sl Phys. Lett.}
{\bf 156B} (1985) 357; {\sl Phys. Rev. Lett}. {\bf 55} (1985) 920.
%
\end{thebibliography}
\end{document}